\documentclass[twocolumn]{aastex63}

\newcommand{\MESA}{{\texttt{MESA}}}

\usepackage{lineno}

\usepackage{amsmath}
\usepackage{enumerate}
\usepackage{graphicx}
\usepackage{hyperref}
\hypersetup{
   colorlinks=true,
    linkcolor=blue,
    filecolor=magenta,      
    urlcolor=blue,
    pdftitle={Overleaf Example},
    pdfpagemode=FullScreen, }
\shorttitle{Unified $g$-mode and $r$-mode Analysis of Accreting White Dwarf Stars}
\shortauthors{Kumar and Townsley}

\graphicspath{{./}{figures/}}

\accepted{to The Astrophysical Journal}

\begin{document}
\title{Unified $g$-mode and $r$-mode Analysis of Accreting White Dwarf Stars}
\correspondingauthor{Praphull Kumar}
\email{pkumar5@crimson.ua.edu, praphullkr05@gmail.com}
\author[0000-0002-8791-3704]{Praphull Kumar}
\author[0000-0002-9538-5948]{Dean M. Townsley}
\affiliation{The University of Alabama,
 Tuscaloosa, Alabama, USA}

\begin{abstract}
Dwarf novae are a subset of cataclysmic variables that accrete material intermittently in short-duration outbursts with sometimes long quiescent intervals in between. During the quiescent state, the white dwarf (WD) photosphere may be observable. Some of these systems show periodic variability consistent with a non-radial oscillation. Asteroseismology has become a unique tool for the measurement of internal structure of the WDs, such as their masses, radii, temperatures, and rotation profiles. A few stable periodicities have been observed for accreting WDs, but the lack of complete and accurate theoretical models has hindered the real diagnosis of the observed pulsations. Though the associated pulsations in accreting WDs are thought to be $g$-modes, some work in the literature suggests that these pulsations could be Rossby modes ($r$-modes). Here, to elucidate this, we present a first simultaneous analysis of $g$- and $r$-mode pulsations in accreting white dwarfs including a full computation of visibility accounting for the distribution of variation over the WD surface. We show that, up to the second lowest degree ($\ell =2$), neither $g-$ nor $r$-modes have a clear advantage in visibility. Although a few retrograde $r$-mode orders exhibit a larger visibility, the low-order $g$ modes possess higher frequency in the star's frame, making them more likely to be driven within the convective driving scenario commonly applied to isolated WDs. Therefore, we favor a $g$-mode origin for the observed periods in accreting WDs, though $r$-modes will be important for stars with more observed modes.

\end{abstract}


\section{Introduction} \label{sec:intro}
More than tens of pulsating accreting white dwarfs (WDs; those in close binaries with mass transfer) have been observed \citep{Szkody_2002, Szkody_2003, Szkody_2004, Szkody_2005, Gansicke_2006, Mukadam_2013, Szkody_2021} since the first discovery of non-radial oscillations in accreting WDs within the cataclysmic variable (CV) GW Lib during quiescent phases before and after an outburst \citep{Warner_1998}. Efforts toward a comprehensive understanding of these systems are still ongoing. Asteroseismology on accreting WDs is uniquely interesting, as it allows a direct probe of how the accretion of mass and angular momentum affects the WD and its subsequent evolution. Such studies can reveal the extent of various compositional layers, the size of the solid core, core temperatures, and more \citep{Sion_1995, Townsley_Bildsten_2004, Godon_2006, Fontaine_2008, Aerts_2010, Romero_2012, Romero_2017}.

Pulsations in isolated WDs consist of gravity modes ($g$-mode), with buoyancy acting as the restoring force. These $g$-modes are excited thermally by the convection zone, whose extent depends on whether the WD surface is primarily hydrogen or helium, as in DAV and DBV stars, respectively (\citealt {Brickhill_1991, Wu_Goldreich_1998, Arras_Townsley_Bildsten_2006, Saio_2013}; subsequent to earlier work by \citealt{Dolez_and_Vauclair_1981, Winget_1982}). However, accreting WDs are distinctive, as they are spun up due to accretion and have surface layers whose chemical composition has not been separated by element diffusion, as the constant addition of accreted materials replenishes these layers \citep{Koester_2009}. 

Two aspects of CV evolution conspire to make non-radial oscillations visible in the accreting WD systems. First, for variations in surface brightness to be visible, the accretion luminosity itself must be less than that from the WD photosphere. This occurs regularly in CVs with low mass transfer rates, typically those below the period gap ($P_{\rm orb}\lesssim 2 h$), which undergo sporadic accretion outbursts observed as dwarf novae. It turns out the typical mass transfer rate for these systems puts the WD surface temperature near the instability strip \citep{Townsley_Bildsten_2004}. Second, the composition of the accreted material, being a roughly solar mixture of hydrogen and helium, shifts the instability strip to slightly higher temperatures than seen for conventional ZZ Ceti stars. For ZZ Ceti stars, which have pure hydrogen atmospheres, empirically the instability strip extends from $\simeq 11000-12600$~K \citep{Gianninas_2011, Tremblay_2015}. The location of the blue edge also depends on the thickness of the hydrogen envelope and the mass of the star \citep{Romero_2013}.

Efforts have been made to investigate this scenario of dislocated instability strip in accreting WDs theoretically, first, \cite{Arras_Townsley_Bildsten_2006} showed that the helium enrichment on the WD surface from the donor will lead to an increase in the temperature of the blue edge of the instability strip compared to the ZZ Ceti strip, though their calculations did not include the mode damping. \cite{Van_Grootel_2015} extended this discussion, but with a non-adiabatic treatment within the time-dependent convection theory, explicitly calculating both the red and blue edges of the instability strip for the GW Lib pulsators across various compositions, and confirmed the predicted shift of the blue to higher surface temperature.

\cite{Szkody_2012} measured the GW Lib rotation period of 209 seconds, comparable to or shorter than the observed oscillation periods of these stars \citep{Vanzyl_2000, Vanzyl_2004}; hence, the rotation treatment is essential and must be done non-perturbatively. Consequently, both gravity and Rossby waves (also known as $r$-mode in an astrophysical context) should be considered in accreting WDs. Rossby modes are the toroidal oscillations of a rotating star for which the Coriolis force participates as a restoring force. They are retrograde-propagating waves of radial vorticity. The possibility of $r$-modes in rotating stars has been long discussed theoretically \citep{Papaloizou_1978, Provost_1981, Saio_1982, Dziembowski_Kosovichev_1987}, but the first clear signature was found only recently via Kepler light curves \citep{Van_Reeth_2016, Li_2019}. 

The potential significance of $r$ modes in accreting WDs was not addressed until recently by \cite{Saio_2019}, who examined their visibility distributions under the assumption of energy equipartition and favored the $r$-mode signatures as the origin of pulsations in accreting WDs. Moreover, these visibility predictions were found to be consistent with the amplitude distributions in $\gamma$ Dor stars. The amplitude distribution of $r$ modes in $\gamma$ Dor stars \citep{Van_Reeth_2016}. However, CV WDs are structurally very different from $\gamma$ Dor stars, and the applicability of these results requires further investigation, which we undertake here.

\cite{Saio_2019} also emphasized the relatively high visibility of $r$-modes compared to $g$-modes, explicitly computing their expected visibility among various order $r$-modes.  Building on our earlier work on $g$-modes in accreting WDs \citep{Kumar_Townsley_2023}, we are motivated here to compute the visibilities of both $g$ and $r$-modes for a representative accreting WD model. Treating the $g$ and $r$ modes with similar methods will allow a more direct comparison. We also consider the role of driving, since the relation between oscillation frequencies in the stellar fluid rest frame (co-rotating) and the observed frequencies is different for $g$ and $r$ modes. Furthermore, the WD models presented in \cite{Kumar_Townsley_2023} were evolved through multiple accretion outbursts but did not include element diffusion, resulting in an unrealistically sharp interface between the envelope and the core. Since the mode frequencies are expected to be sensitive to the structure of the core-envelope boundary, we include the element diffusion during the long-term accretion phase of the WD.

This paper aims to compute and show the visibilities for both gravity and Rossby modes in accreting WDs. First, section \ref{sec:wdmodel} supplies information on the thermal structure of the WD and elucidates constructing WD models informed by expected observational constraints. Section \ref{ssec:soltolte} describes the solutions of the Laplace Tidal equation and discusses the property of the eigenvalues. The Visibility computations and their relevance to $g $ and $r$ modes are elucidated in section \ref{ssec:visibilitycomputation}. The validity of the TAR is presented in section \ref{sec:tarvalidation}. Section \ref{sec: discussionandconclusions} concludes the paper and discusses planned future work.


\section{ White Dwarf Thermal Structure} \label{sec:wdmodel}
The sensitivity of normal modes of a WD to its interior structure necessitates a careful consideration of the theoretical models employed. The long-standing uncertainty concerning the mechanism of angular momentum loss from CV binaries severely limits our observational and theoretical understanding of the evolution of CVs \citep{Nelemans_2016, Schreiber_Zotorovic_Wijnen_2016}. CV binaries are believed to lose angular momentum via various mechanisms, such as magnetic braking or gravitational wave radiation. This loss,
and the resulting binary evolution, dictates the accretion rate onto WD \citep{Howell_2001, Knigge_2006, Knigge_2011}. WDs with a longer orbital period $P_{\rm orb} \ge 3 h$ exhibit a higher mass accretion rate from their donors as compared to shorter orbital periods with $P_{\rm orb}<2h$ (below the period gap). Direct measurement of angular momentum loss is difficult, thus $T_\mathrm{{eff}}$ serves a good measure of angular momentum loss determined by the compressional heating of the accreted material \citep{Townsley_Bildsten_2004, Townsley_and_Gansicke_2009}, providing constraints on the mean accretion rate, $\langle \Dot{M} \rangle$, of the WDs. The WDs with shorter orbital periods can potentially accumulate  $\sim 10^{-10}~\mathrm{M_\odot yr^{-1}}$ from their companions \citep{Kolb_Baraffe_1999, Townsley_Bildsten_2003, Pala_2017, Pala_2022}. This accretion rate can potentially be up to $\sim 10^{-9}~\mathrm{M_\odot yr^{-1}}$ for CVs with an orbital period above the period gap, resulting in higher observed $T_{\text {eff}}$ \citep{Howell_2001}.

\begin{figure}
    \centering
    \includegraphics[width=1.\linewidth]{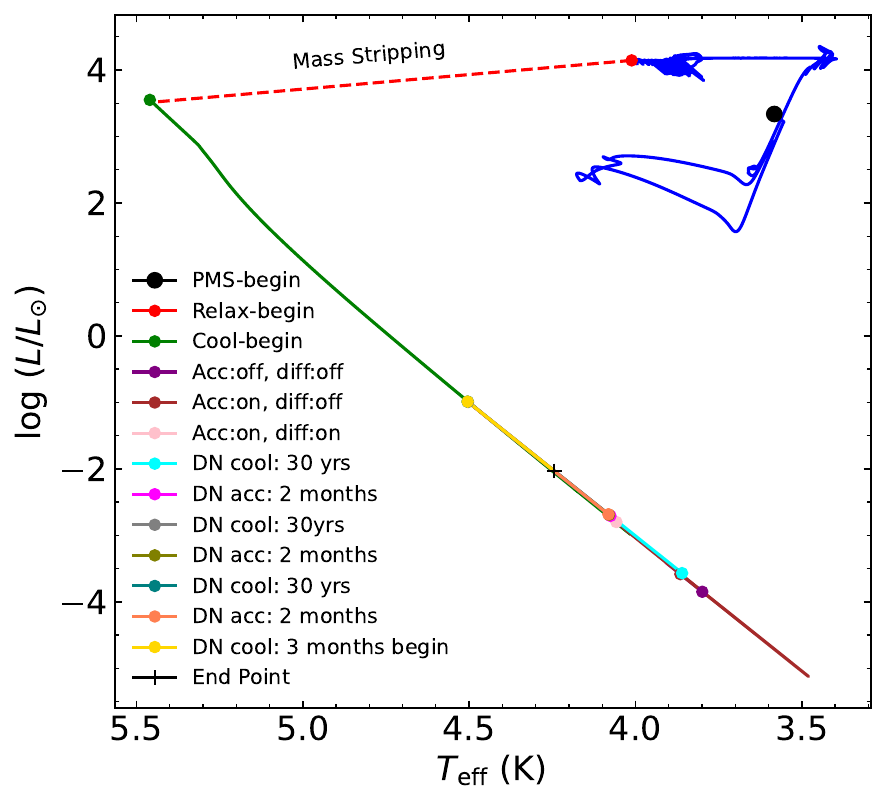}
    \includegraphics[width=1.\linewidth]{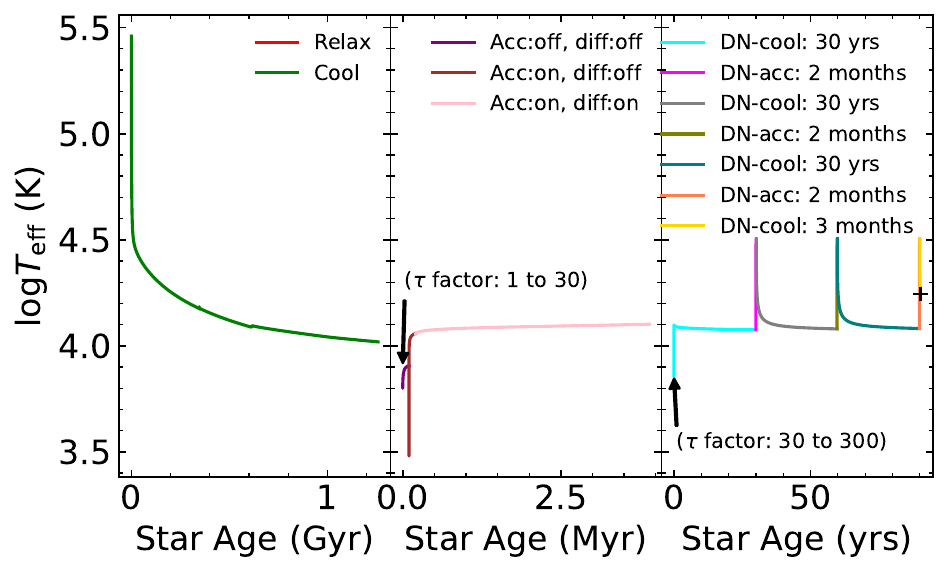}
    \caption{HR diagram (top) and effective temperature evolution (bottom) of a 0.78~$M_\odot$ WD through its evolutionary stages from pre-main-sequence (PMS) to cooling. Tracks highlight accretion (on/off) phases with element diffusion and dwarf nova (DN) cycles. The sharp drops in $T_{\rm eff}$ in the second and third bottom panels (black arrows) correspond to changes in boundary conditions, specifically the adopted optical depth ($\tau$) at the outermost zone. We perform our seismological analysis on the final model, marked with a '+'.}
    \label{fig:hrdiagram}
\end{figure}

\subsection{Previous Work and Limitations} \label{ssec:previouswork}
We build upon the earlier work of \cite{Kumar_Townsley_2023}, who presented WD models for accreting systems that undergoing multiple nova outbursts. Their WDs were prepared in a way similar to that used in \cite{Timmes_2018}. We briefly summarize one of the WD models from \cite{Kumar_Townsley_2023}, which we adopt in this study. Their model evolves a pre-main sequence star ($M_{\rm initial} = $3.95~$M_\odot$) to a hot WD using one-dimensional stellar evolution code in \MESA\ version 10398 \citep{Paxton_2011, Paxton_2013, Paxton_2015, Paxton_2018, Paxton_2019}, with solar metallicity and adopting OPAL opacity tables \citep{Rogers_Nayfonov_2002}. Convection is treated using mixing length theory in the `\textit{Henyey}' formulation, with a mixing-length parameter $\alpha_{\rm MLT}=1.9$. After the remaining hydrogen from the surface is stripped off, the final 0.78~$M_\odot$ WD is then cooled including the element diffusion, until its effective temperature reaches 15,000~K. The $T_{\rm c}$ of this WD corresponds to $1.26\times10^7$~K, and we use this as a starting model for our study and subsequent evolution.

For an observed accretion rate, these CVs (below the period gap) should possess a much lower $T_c$, so that WD surface temperature matches the observed $T_{\mathrm{eff}}$ \citep{Townsley_Bildsten_2004}. 
Ideally, we would be able to simulate the evolution of $T_{\rm c}$ as the WD undergoes many H shell flashes, with their accompanying classical nova ejection events. We were not able to accomplish such for this work. Due to our inclusion of element diffusion (described next), convective boundary mixing at the base of the H-rich layer during the H flash is expected to play a critical role in determining how the flash proceeds. Notably, the amount of material that is dredged up, and the amount left on the surface after ejection are important quantities needed in order to include H shell flashes in the energy budget. However, we were not able to obtain tractable numerical behavior of the convective boundary during the runaway with the convective boundary methods available in the \texttt{MESA} versions we have used so far. We consider this mixing during the runaway to be an active area of current and future research \citep{Jose_2020}, which successful seismology might be able to inform. As such, we have taken $T_{\rm c}$ as a parameter that we will vary within some range motivated by the considerations of how thermal equilibrium may be approached over many flashes as discussed by \cite{Townsley_Bildsten_2004}.

\begin{figure}
  \includegraphics[width=1\linewidth]{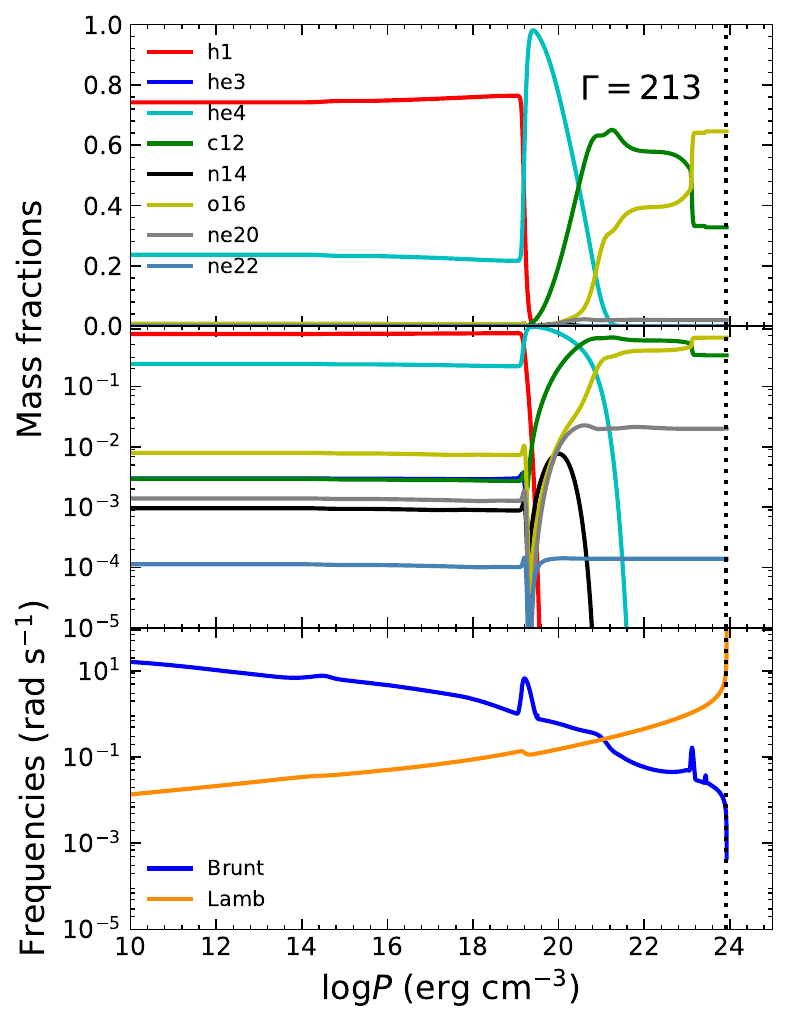}
   \caption{Displayed are the various elemental abundance profiles in the top two panels in linear and logarithmic scale plotted against the pressure for the 0.78~$M_{\odot}$ model, three months after the dwarf nova outburst (short-term accretion). Brunt-V\"ais\"al\"a and Lamb frequency are shown in the bottom panel. Mass fractions are shown on the log scale in the middle panel to highlight the other species produced, which are difficult to note in the linear scale.}
    \label{fig:brunt abundance for 0.78 Msun} 
\end{figure}

\subsection{This Work} \label{ssec:thiswork}
In this work, we continue to move toward a more realistic WD model incorporating more physics and better characterization of evolution. Compared to \cite{Kumar_Townsley_2023}, we now include element diffusion during the accretion process, instead of only during the WD cooling. This is essential for having a physically valid structure for the core-shell boundary, which influences the mode frequency spacings. We adopt an accretion rate more consistent with that expected for the stage of binary evolution the CV WD systems are in \citep{Townsley_and_Gansicke_2009, Knigge_2011}, which also requires a lower $T_{\rm c}$ in order to match observed effective temperatures \citep{Townsley_Bildsten_2004}. Due to encountering apparently intractable numerical challenges in simulating multiple hydrogen shell flashes with convective boundary mixing, we have elected to consider accretion onto a bare WD and evaluate the seismological properties during accretion up to the onset of the hydrogen shell flash. We find this alternative preferable to an unrealistic, numerically problematic, and unconstrained mixed layer. While the precise evolutionary pathways of accreting WDs remain unclear \citep{Schreiber_Zotorovic_Wijnen_2016, Shen_2022}, we incorporate the long-term effects of accretion on the WD interior structure assuming constant accretion rate over extended timescales.

We utilize a 0.78~$M_\odot$ WD model from \cite{Kumar_Townsley_2023}, starting our evolution for this work when the WD reaches $T_{\rm c}=1.26\times10^7$~K. However, for the subsequent evolution, we switch to more recent version of \texttt{MESA v15140}, which offers more stable and robust timestep controls. The WD is then cooled to $T_{\rm c} = 5\times 10^{6}$~K with element diffusion enabled, resulting into a smooth abundance profile at the base of the surface He layer. Following this cooling stage, the WD is subject to long-term accretion with a time-averaged rate $\dot M= 6\times 10^{-11}~M_\odot ~\mathrm{yr^{-1}}$ (initial $\mathrm{Y} = 0.237$ and $\mathrm{Z} = 0.02$) using \texttt{pp\_cno\_extras\_o18\_ne22.net}, with both element diffusion and thermohaline mixing (unity mixing coefficient) included. The different evolutionary stages of the WD are shown in the top panel of Figure \ref{fig:hrdiagram}. The accreted material is added uniformly onto the WD surface with the thermal state of the photosphere under the assumption of spherical symmetry \citep{Paxton_2015, Townsley_Bildsten_2004}. 

The core temperature of the WD is approximately $5\times10^6$~K, which is relatively low and typical for CVs with the observed $T_{\rm eff}$ \citep{Townsley_Bildsten_2004}. At this $T_{\rm c}$, the WD core is approximately 8.4\% crystallized. Based on phase diagrams presented in recent studies and core compositions of our model ($X(^{12}$C) = 32.76\% and $X(^{16}$O) = 64.57\%), we adopt the beginning of the crystallization at a plasma Coulomb parameter of $\Gamma > 213$ \citep{Blouin_2021, Bedard_2024}. Notably, this differs from the canonical one-component value of $\Gamma > 175$ \citep{Segretain_1993, Isern_1997, Potekhin_chabrier_2000, Potekhin_chabrier_2010}. However, this is still an active area of research. Choosing a higher $\Gamma$ pushes the crystallization transition inward, thereby changing the individual mode frequencies by a few percent, though the overall mode structure remains largely unaffected. We do not explicitly evaluate how mode frequencies may vary due to changes in the extent of crystallization, we will explore different sizes of the crystallized core in detail in future work. 

Additionally, a small mass fraction of $^3$He ($\approx 3\times 10^{-3}$) is included in the accreted material, compared to the $\sim 10^{-3}$ expected for realistic donors, leading to more pre-ignition heating due to nuclear burning. Due to the numerical difficulty of \texttt{MESA} during the mass-loss phase with the overshooting (convective dredge-up) and diffusion, we halt the evolution at two-thirds of the time ($\sim 3.91$ million years) that it takes to reach the first hydrogen flash. This corresponds to the hydrogen luminosity $\log(L/L_\odot)=-3.06$. This numerical difficulty with overshoot means that we are unable to evolve the model through multiple nova outbursts. This effectively leaves us with a choice between two models for the WD structure: (1) models that have been evolved through multiple nova outbursts, but without diffusion giving an unrealistically sharp interface between envelope and core like those in \cite{Kumar_Townsley_2023}, or (2) models that have a realistic boundary due to having element diffusion included, but have not undergone multiple nova outbursts. We choose to present (2) here, as we believe it is more important to have a realistic core envelope interface at this point in the development of seismological theory for these objects. The mode frequencies are expected to be sensitive to the structure of this core-envelope boundary. Improvements to the evolutionary model can be revisited in future work. The WD's final radius and $T_{\rm eff}$ at the end of long-term accretion are 0.0104~$R_\odot$ and  12360~K, respectively. The surface gravity of the WD $\log g \approx 8.29$. 

The thermal instability of the accretion disk causes the donor to transfer mass rapidly in intermittent bursts (dwarf novae). In order to capture this, following the long-term accretion phase, we transition the modeling to the dwarf nova cycle. During this phase, the WD undergoes episodic accretion at a rate of $\dot{M}=1.2\times 10^{-8}$~$M_\odot$~yr$^{-1}$ for a short duration of 2 months, with a recurrence time of 30 years. The bottom panels of Figure \ref{fig:hrdiagram} show the evolution of the surface temperature from the initial mass relaxation phase to subsequent dwarf nova cycles. We adopt a slightly higher $\tau$ in the outermost zone to ensure the numerical stability during the accretion phase (indicated by black arrows), which results in sharp drop of $T_{\rm eff}$. 

Although rotation is included throughout the cooling phase of the WD, we do not include it during the accretion phase due to numerical difficulties. Instead, rotation has been added explicitly in \texttt{GYRE} \citep{Townsend_2013, Townsend_2018} when computing the mode frequencies of the star. However, we expect that this adjustment may change the mode frequencies of the both $g$ and $r$ modes by a few percent. We will address it in our future work. For this study, we compute the mode frequencies three months after the short-term accretion event using \texttt{GYRE 7.0} version. The relevant input files and data sets are available online at doi: \url{https://doi.org/10.5281/zenodo.17765239}.

Figure \ref{fig:brunt abundance for 0.78 Msun} shows the WD compositions, buoyancy and Lamb profiles three months after the short-accretion event of a 0.78~$M_{\odot}$ model. To highlight the contributions from the minor species, we plot the mass fractions in both the linear (top panel) and logarithmical (middle panel) scales. The smooth composition profiles are the result of the use of element diffusion during the long-term accretion phase. The peaks presented in the Brunt-V\"ais\"al\"a frequency correspond to the composition gradients produced by the elemental species. The peak in buoyancy frequency $\log P \approx 19$ $\mathrm{erg ~cm^{-3}}$ reflects the transition between the hydrogen and helium layers. 

Both the larger and the smaller peaks in the Brunt-V\"ais\"al\"a profile between $\log P$ of 22 and 24~$\mathrm{erg ~cm^{-3}}$ are contributed by the CO gradient left at the outer edge of the remnant of the convective core formed during the core He burning phase.

\section{Joint Analysis of $g$ and $r$ Modes} \label{sec:grmodes}
In this section, we present the solution to the full equation of motion, detailing the methods employed during this course to obtain the eigenfrequencies of both $g$ and $r$ eigenmodes. We also provide a modest overview of how the mathematical structure of the eigenspace leads to the relevant mode branches, incorporating conventions from previous literature. Furthermore, by computing the visibility function on the star's surface, we illuminate the relevance of both $g$ and $r$ modes observed in the accreting WD.
\subsection{Solutions to Laplace's Tidal Equation (LTE)} \label{ssec:soltolte}
\begin{figure}
    \includegraphics[width=1\linewidth]{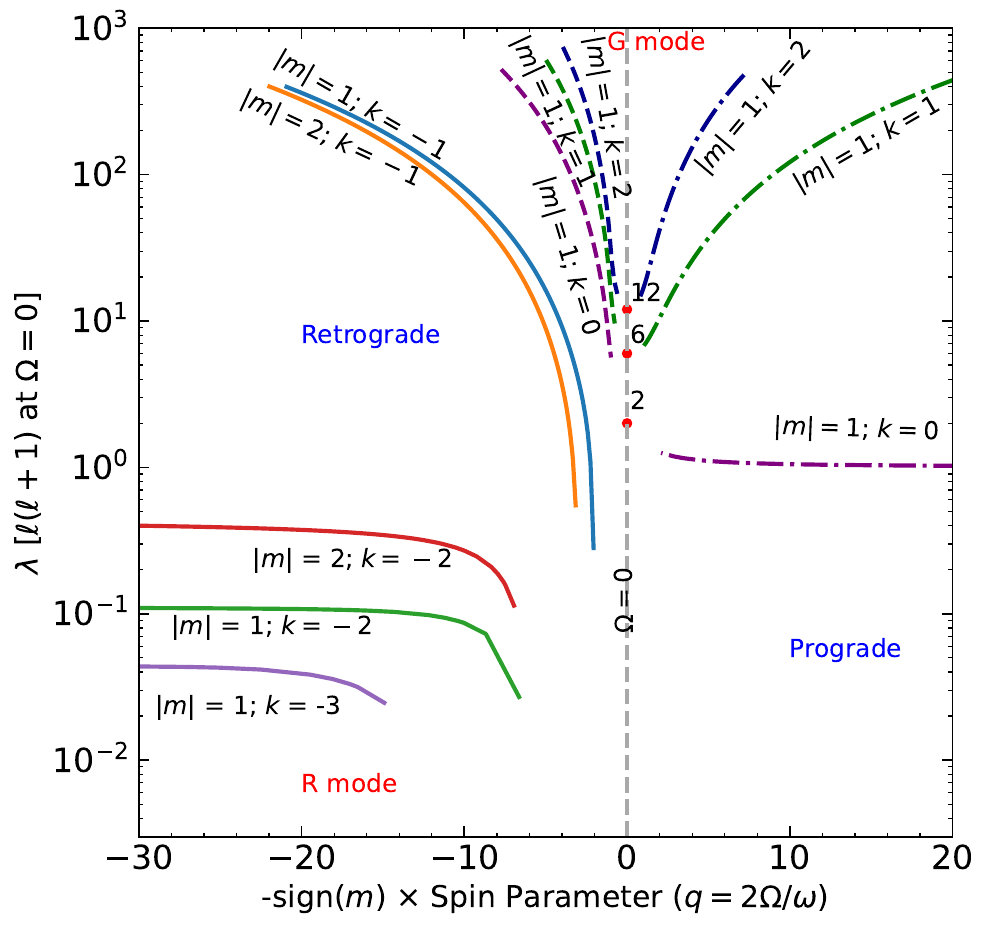}
    \caption{Eigenvalue ($\lambda$) of Laplace's Tidal equation (\ref{eqn:LTE}) plotted against the spin parameter ($q = \frac{2\Omega}{\omega}$). $\Omega$ is the angular rotation frequency of the star, and $\omega$ is the mode frequency in the star's frame. Eigenvalues of both $r$ (Rossby)- and $g$ (gravity)-modes are displayed in this plot. The gray dashed vertical axis is the zero rotation ($\Omega = 0$) line. Modes with $m < 0$ ($m > 0$) propagate in the prograde (retrograde) direction in the co-rotating frame, respectively, and are marked on the right (left) side of the panel. Solid lines correspond to $r$ modes, and dashed (dot-dashed) lines indicate retrograde (prograde) $g$ modes, respectively. At zero rotation, $\lambda$ approaches $\ell(\ell + 1)$, which is the solution of LTE in the non-rotating limit indicated as red dots and marked with green text. The prograde and retrograde mode lines don't connect at zero rotation because the values shown are for actual modes in a finite rotation frequency star, thus having a maximum $\omega$, and corresponding minimum $q$.}
    \label{fig:lambdaspin} 
\end{figure}

\begin{figure}
  \includegraphics[width=0.92\linewidth]{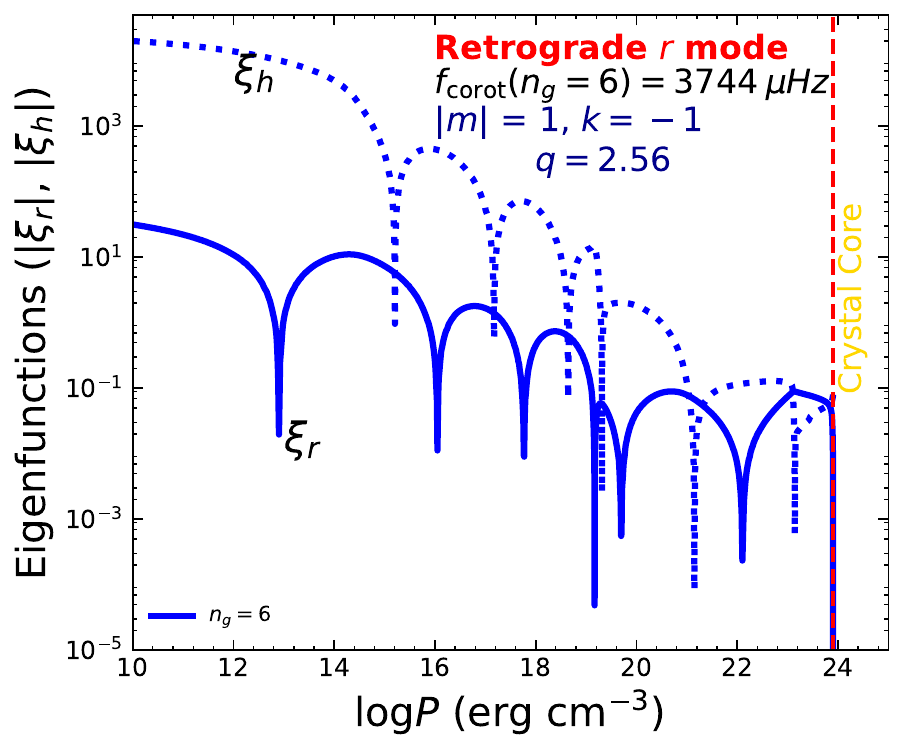}
  \includegraphics[width=0.92\linewidth]{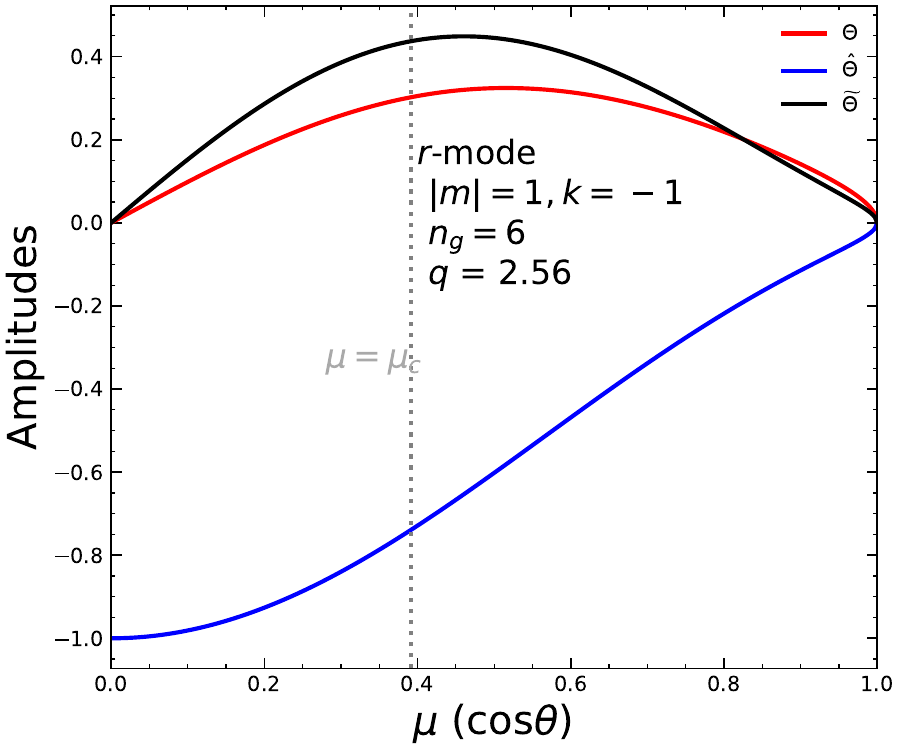}
  \includegraphics[width=0.92\linewidth]{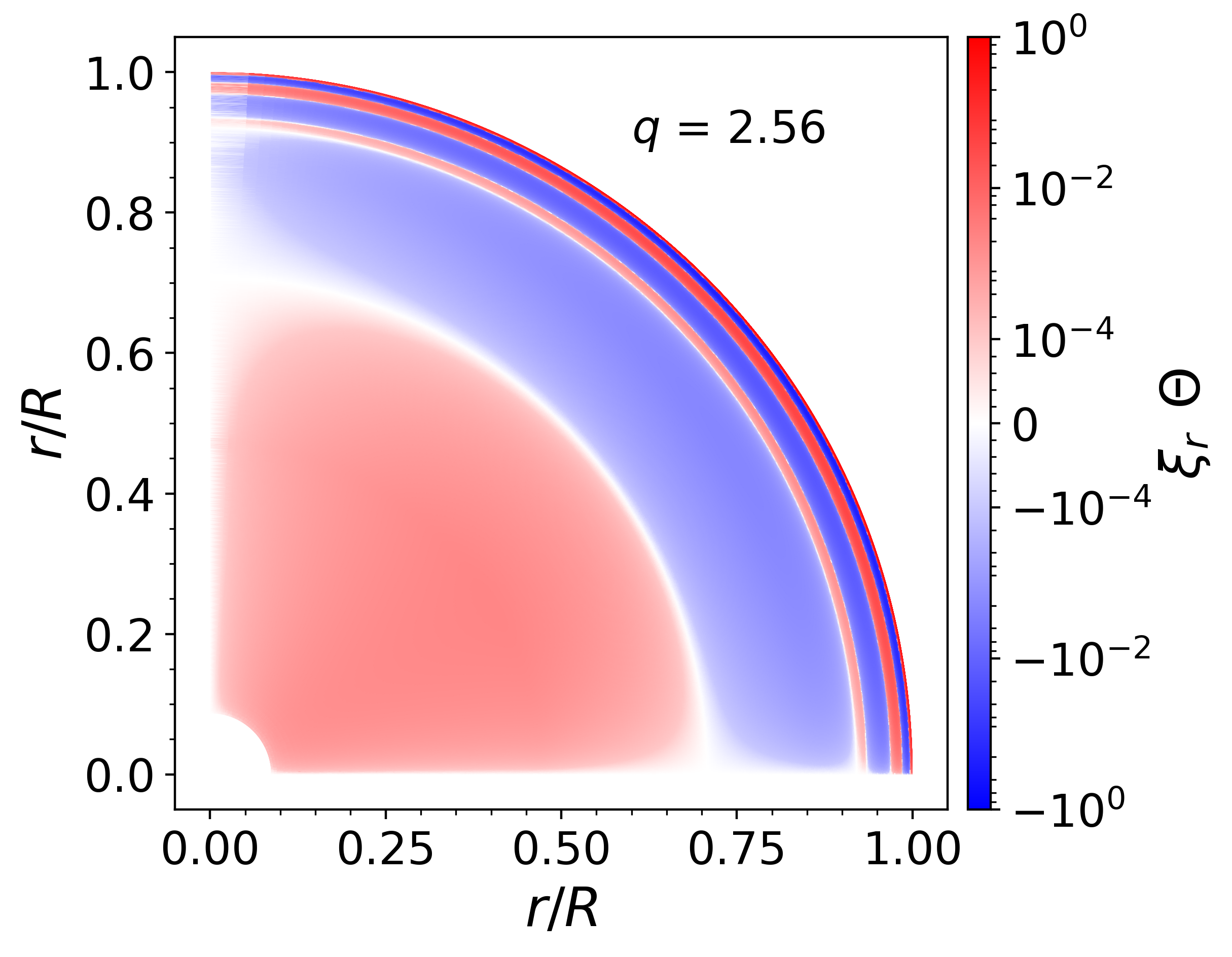}
   \caption{Absolute radial and horizontal displacement perturbations of the sixth radial order of, $m = 1$, $k = -1$, $r$ mode are shown in the top panel. The WD core is to the left. The dashed red line indicates that 8.4\% of the WD core is crystallized. The corresponding unnormalized angular eigenfunctions (also known as Hough functions), which are the solutions of LTE, with spin parameter $q = 2\Omega$/$\omega$ = 2.56, $\Theta, \hat{\Theta}$, and $\Tilde{\Theta}$, are shown in the middle panel. The vertical dashed line indicates the boundary of the equatorial waveguide, the critical angle ($\mu_c$), $\mu = 1/q$. The bottom panel exhibits the dependence of the full displacement
   eigenfunction, $\xi_r(r)\Theta(\theta)$, on both $r/R$ and $\theta$ in the meridional plane.}
    \label{fig:hough and radial eigenfunctions} 
\end{figure}
A modestly more detailed discussion of how the mode equations arise can be found in our previous work \citep{Kumar_Townsley_2023}, where the emphasis was on $g$ modes under the rapid rotation, or, for more complete coverage, any of several excellent references in the literature \citep{Lee_Saio_1997, Unno_1989, Townsend_2003a, Bildsten_Ushomirsky_Cutler_1996}.
Several approximations are made to obtain the mode equations that we will solve for both the $g$ and $r$ mode families. For a uniformly rotating star, within the Cowling approximation \citep{Cowling_1941}, where the Eulerian perturbations of gravitational potential are neglected and the solutions are assumed to be oscillatory ($e^{i\omega t},~\omega$ is the mode oscillation frequency in the corotating frame), the linearized momentum equation in the star's frame is given by \citep{Bildsten_Ushomirsky_Cutler_1996, Townsend_2003a},
\begin{eqnarray} \label{eq:momentum}
i\omega\frac{d\vec{\xi}}{dt} &=& -\frac{\nabla P}{\rho} - g \hat{r} - 2i\omega\vec{\Omega} \times \vec{\xi}
\end{eqnarray}
where $\vec{\xi}$ is the Eulerian displacement, and $\rho$ and $P$ are the density and pressure of the unperturbed stellar structure, respectively, and $\vec{g}$ is the gravitational acceleration. The final term corresponds to the Coriolis force. 

Observations suggest that $g$-mode period spacing is strongly influenced by rotation \citep{Bouabid_2013, Van_Reeth_2015}. Therefore, a complete 2-dimensional treatment is optimal. Due to both analytical difficulty and the computationally expensive nature of the problem, we continue, in this work, to work within the ``Traditional Approximation for Rotation" (TAR) in the equation of motion to obtain the eigenmodes and eigenfunctions of both $g$ and $r$ modes. This allows the system to become separable into an angular and radial part of the eigenfunctions \citep{Eckart_1961, Lee_Saio_1987, Lee_Saio_1997, Bildsten_Ushomirsky_Cutler_1996, Townsend_2005}.

Within this approximation, the horizontal component of the stellar rotation vector ($\Omega\sin\theta$, $\Omega$ is the rotation frequency of the star and $\theta$ is the co-latitude) in the perturbed momentum equations is neglected \citep{Kumar_Townsley_2023} (assuming uniform rotation). Furthermore, TAR greatly simplifies the understanding of Coriolis force on low-frequency oscillations, though also provides a good estimate for high-frequency modes \citep{Lai_1997}. The accuracy and reliability of TAR is still unclear. Moreover, this is a good approximation in the outer part of the star and not so near the central regime of the star. We anticipate that the approximation may be accurate for the entire star if a significant fraction of the WD core is crystallized. We consider here where this may or may not be the case, finding that much of the parameter space relevant to observed objects is unlikely to have a sufficiently large solid core for the TAR to be entirely sufficient.

Following the TAR application in equation (\ref{eq:momentum}) and assuming the temporal dependency, the separated radial perturbation can be expressed as,
\begin{eqnarray} \label{eqn:oscillationform}
\xi_r \left(r, \theta, \phi \right) &=& \xi_{r} (r) \Theta (\theta ; q) e^{i\omega t + im\phi}
\end{eqnarray}
where $\xi_{r} (r)$ is the radial displacement amplitude, $\Theta$ is the Hough function \citep{Hough_1898}, and $m$ is the azimuthal order. Within these conditions, the governing equation is reduced to the same equation for the non-radial pulsation of a non-rotating star, except that $\ell(\ell + 1)$ is replaced with $\lambda$, which is the eigenvalue of Laplace's tidal equation, and given by \citep{Bildsten_Ushomirsky_Cutler_1996, Lee_Saio_1997, Kumar_Townsley_2023}:
\begin{eqnarray}\label{eqn:LTE}
\nonumber
\left[ \frac{\partial}{\partial\mu}\left(\frac{1-\mu^2}{1-q^2}\frac{\partial}{\partial\mu}\right) - {} \right. &&\\ \left. \frac{m^2}{(1-\mu^2)(1-q^2\mu^2)} - \frac{qm(1+q^2\mu^2)}{(1-q^2\mu^2)^2}\right]\Theta &=& -\lambda \Theta
\end{eqnarray}
with $\mu=\cos\theta$. In the limit of zero rotation, $\Theta(\theta) e^{im\phi}$ is replaced by $Y_\ell^m(\theta, \phi)$ in the full solutions for a harmonic degree $\ell$ and azimuthal order $m$ \citep{Lee_Saio_1997, Kumar_Townsley_2023}. The eigenvalue $\lambda$ can be visualized as a transverse wave number (i.e., $\mathrm{k}_{tr}^2 = \lambda/\mathrm{R}^2$; $\mathrm{R}$ is the radius of the star). The solutions of Laplace's tidal equation depend on the spin parameter $q=2\Omega/\omega$, which characterizes the relative strength of Coriolis and buoyancy. The importance of the Coriolis force becomes significant for $q \ge 1$. Although the radial and angular differential equations are separable within the TAR, the appearance of $\omega$ in $q$ and, therefore, in the angular eigenvalue equation for $\lambda$, along with the aforementioned appearance of $\lambda$ in the radial equations, means that, for a given $\Omega$, the two eigenvalue problems must still be solved simultaneously for each mode. 

In general, equation \eqref{eqn:LTE} allows an infinite number of eigenvalues and corresponding eigenfunctions. These are sometimes designated as $\lambda_{km}$ and $\Theta_{km}(\mu; q)$, respectively, where $k$ serves as an ordering index, and we will continue to use this notation throughout this paper. The eigenvalue $\lambda_{km}$ implicitly depends on the spin parameter $q$, which itself contains the star's pulsation and rotation frequencies. Physically, $\lambda$ is proportional to the strength of horizontal compression in the positive $\delta p$ phase. When the effect of rotation is small ($q<1$), the oscillations are associated with $p$-mode and $g$-mode character. Although, for $q>1$, when the eigenvalues $0< \lambda \le 1$, the oscillations are associated with $r$-modes. For the negative solutions of $\lambda$, an alternative set of modes is not discussed in this article. We direct readers to follow \cite{Lee_Saio_1997} for more discussion.

We use appropriate boundary conditions to obtain the correct solutions of equation \eqref{eqn:LTE}. We separate two categories of solutions based on behavior at the equatorial plane, $\mu=0$. That is, $\Theta(-\mu) = \Theta(\mu)$ for an even solution and $\Theta(-\mu) = -\Theta(\mu)$ for an odd solution. These solutions are obtained separately by applying different boundary conditions at $\mu = 0$, i.e., $\frac{d\Theta}{d\mu} = 0$ for an even solution and $\Theta = 0$ for the odd solution. We use a series method combined with shooting to obtain the final eigenvalues and eigenfunctions (see \cite{Bildsten_Ushomirsky_Cutler_1996} for further discussion). The oscillations are proportional to $e^{i(\omega t + m\phi)}$, and a prograde (retrograde) has a negative (positive) value of $m$. 

Two families of eigenvalues exist, one for which any value of $q$ exists and the other exists only when $|q| > 1$. These are categorized with $k \ge 0$ and $k \le 0$. The eigenvalues with $k \ge 0$ are given by $\lambda_{km} = (|m| + k)(|m| + k + 1)$ at $q = 0$, which reduces to $\ell(\ell + 1)$ for $\ell = |m| + k$ \citep{Lee_Saio_1997}. This is the solution for a non-rotating star. Although there is no such universal nomenclature, in this work we adopt the convention that eigenvalues with $k \ge 0$ correspond to $g$-mode oscillations, while those with $k < 0$ correspond to $r$-mode oscillations, which only exist in the presence of a finite rotation. 

For $k < 0$ and $m > 0$ (retrograde), the oscillations correspond to $r$-mode and only exist with $q > (|m| + k)(|m| + k - 1) $. The frequencies of $r$-modes are bounded in the co-rotating frame \citep{Saio_2018,Saio_2018_conf} as,
\begin{eqnarray} \label{eqn:rmodefreqrange}
\omega \:(r\text{-mode}) < \frac{2m\Omega}{(|m| + |k|)(|m| + |k| - 1)} \le \Omega
\end{eqnarray}
where $\omega$ is the angular frequency of the mode. R-mode propagates in a stably stratified layer, and their small values of $\lambda$ indicate that the displacement is predominantly toroidal. However, $\lambda$ for $k = -1$ increases rapidly as the spin parameter, ${q}$, increases. $\lambda$ for $k \le -2$ continues to remain small even for large $q$ as \citep{Townsend_2003a}:
\begin{eqnarray} \label{eqn:assymptoticlambda}
    \lambda \: \left(q; k \le -2\right) \simeq \frac{m^2}{(2 |k| - 1)^2} \quad \quad \text {if $q \gg 1$}
\end{eqnarray}

Since the overall LTE remains the same for both $g$ and $r$-modes, the frequency ($\nu_{\rm co}$) of a high order ($n_\text{g} \gg 1$) $g$-mode or $r$-mode in the star's co-rotating frame is given as:
\begin{eqnarray} \label{eqn:cyclicfreq}
  \nu_{co} \:(g; r)  = \frac{\sqrt{\lambda}}{n_{g}}\nu_{co,0}
\end{eqnarray}
where $\nu_{co, 0}$ is the fundamental co-rotating frequency.

Figure \ref{fig:lambdaspin} presents the final eigenvalue ($\lambda$) solutions of the Laplace Tidal Equation (\ref{eqn:LTE}) for modes obtained from \texttt{GYRE}, plotted against the spin parameter. Retrograde modes are shown on the left side of the plot. We adopt a spin period of approximately 209 seconds to compute the mode frequencies, which is close to the value observed for GW Lib \citep{Szkody_2012}. The eigenvalues corresponding to $g$- and $r$-modes are displayed. The gray dashed vertical line denotes the zero rotation ($\Omega = 0$), assuming that the star is in the inertial state. The solid lines represent solutions for $r$-modes, which only exist for $q>1$ in the retrograde direction of the star. The frequencies of these modes depend on the rotation frequency of the star and are bounded by equation \ref{eqn:rmodefreqrange}. The dashed lines (on the left side of the figure) indicate the retrograde $g$ modes, while the dot-dashed lines (on the right side) are the prograde $g$ modes. The retrograde and prograde $g$ mode lines merge to $\ell(\ell + 1)$ at $\Omega = 0$, denoted by red dots. Odd and even $k$'s (including $k=0$) demonstrate the odd and even order modes. For $k = 0, -1$, $\lambda$ increases rapidly with the increasing spin parameter for $r$ modes. Although, for $k \le 2 $, $\lambda$ remains small and approaches a constant value at high $q$, as shown in figure \ref{fig:lambdaspin} (cf. equation \ref{eqn:assymptoticlambda}). This behavior has a direct influence on the structure of the surface eigenfunctions, which will be shown later in this manuscript.

There is some ambiguity in mode classification due to mismatches in behavior at small and large $q$ or in the retrograde and prograde directions. As a result, authors have made slightly different choices about how to index and classify modes \citep{Lee_Saio_1997, Townsend_2003a}. Table \ref{table:table_mode_classification} lists the orders of the $g$ and $r$ modes explored and discussed in this work. We use a $k$ indexing scheme similar to \cite{Townsend_2003a}, capable of uniquely identifying all modes. Our primary focus is on the low-order modes with periods roughly between 100 and 1000 s. Gravity modes have well-defined $\ell$ values (see 1st column of table \ref{table:table_mode_classification}), unlike Rossby modes with toroidal wave functions. Note that \texttt{GYRE}, in its parameter files, uses an $\ell$ scheme rather than the $k$ scheme; it sets $\ell = |k| +|m| -1$ for the Rossby mode. The $\ell = -m$ or $k = 0$, prograde sectoral mode, which characterizes gravity wave properties, is identified as the Kelvin mode \citep{Gill1982AtmosphereOceanD}. $|k|=1$ mode possesses special behavior, the $k=-1, m=1$ acts like an $r$-mode in the retrograde direction, and $k=1, m=-1$ behaves like $g$-mode in the prograde direction; this is first described by \cite{Yanai_Maruyama_1966}, namely Yanai mode. The $m=0$ corresponds to zonal mode.

The $\theta$-dependence of the displacement vector, i.e., the  horizontal displacement vector, can be given similarly to equation \ref{eqn:oscillationform}:
\begin{eqnarray} \label{eqn:thetadependence}
\left(\xi_{\theta}, \xi_{\phi}\right) &=& \left(\frac{\xi_{h} (r) \hat{\Theta} (\theta ; q)}{\sin\theta}, \frac{\xi_{h} (r) \tilde{{\Theta}} (\theta ; q)}{i\:\sin\theta}\right) e^{i(m\phi + \omega t)}
\end{eqnarray}
where $\xi_{h}(r)$ is the horizontal part of the displacements and $\hat{\Theta}$ and $\Tilde{\Theta}$ are the angular eigenfunctions along the $\theta$ and $\phi$ directions, respectively. The solutions for $\hat{\Theta}$ and $\Tilde{\Theta}$ are given as \citep{Lee_Saio_1997, Townsend_2003a, Townsend_2020}:
\begin{eqnarray}
\label{eqn:houghhat}
    \left[(1-\mu^2) \frac{\rm d}{\rm d\mu} - mq\mu \right] \Theta &=& (q^2\mu^2 - 1)\hat{\Theta}\ , \\
\label{eqn:houghtilde}
    \tilde{\Theta} &=& -m\Theta - q \mu \hat{\Theta}\ .
\end{eqnarray}
Figure \ref{fig:hough and radial eigenfunctions} shows all components of an example eigenfunction, the $r$ mode with $k=-1$, $|m|=1$, and radial order $n_g=6$. This mode has a frequency in the corotating frame of 3744 $\mu$Hz, and thus $q=2.56$. The middle panel displays eigenfunctions used to construct the displacement along the $r$, $\theta$, and $\phi$ directions labeled as $\Theta$ (red line), $\hat{\Theta}$ (blue line), and $\Tilde{\Theta}$ (black line), respectively. These are the direct solutions of equations \ref{eqn:LTE}, \ref{eqn:houghhat}, and \ref{eqn:houghtilde}. The amplitudes of the Hough functions are scaled similaraly to \cite{Lee_Saio_1997} - odd $k$'s are scaled such that $\Theta (\mu = 0) = 0$ and $\hat{\Theta}(\mu = 0) = -1$ and even $k$'s are scaled so that $\Theta (\mu = 0) = 1$, and $\hat{\Theta} (\mu = 0) = 0$. 

\begin{figure}[h]
    \includegraphics[width=1\linewidth]{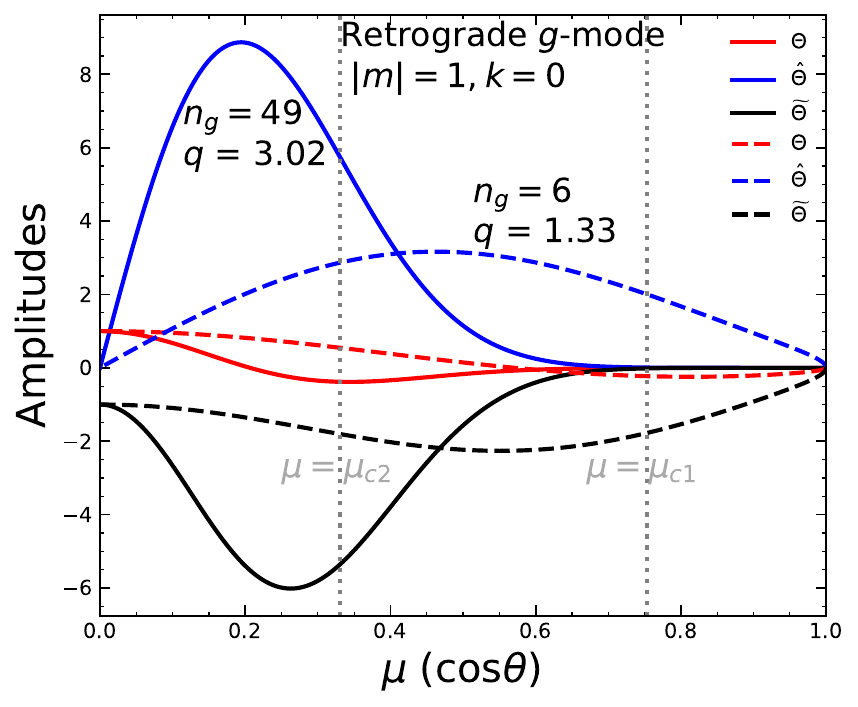}
    \includegraphics[width=1\linewidth]{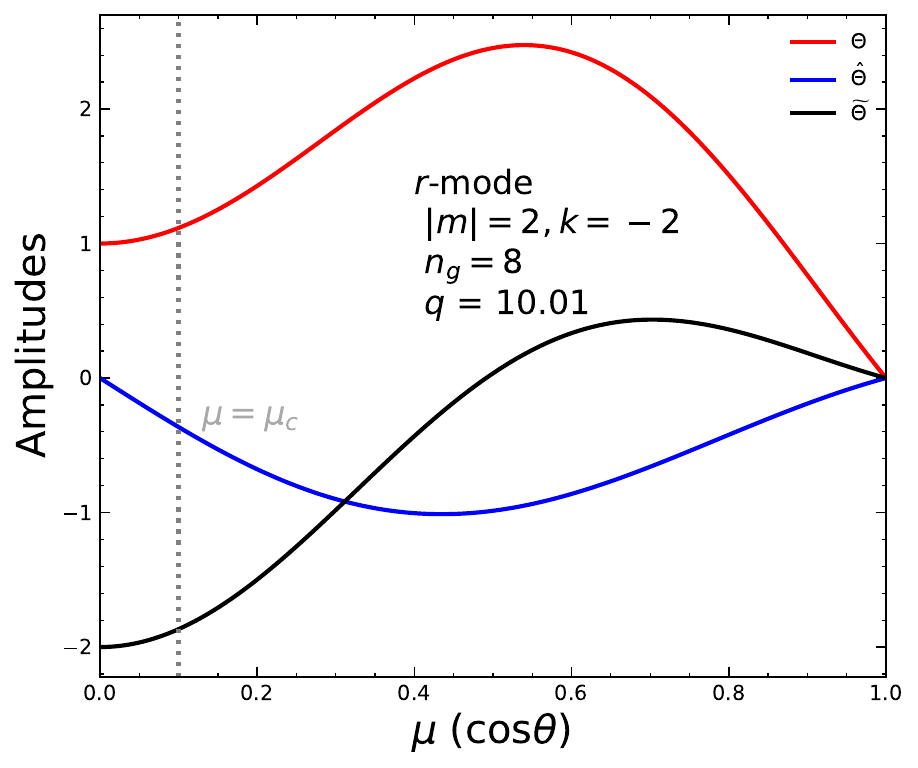}
    \caption{ Unnormalized but scaled (explained in the text) Hough eigenfunctions $\Theta$, $\hat{\Theta}$, and $\Tilde{\Theta}$ are plotted as functions of $\mu =\mathrm{cos \theta}$ for the lowest order retrograde $g$ mode (top panel) with $|m| =1, k=0$, where solid lines are for $q = 3.02$ and dashed lines are for $q = 1.33$. The bottom panel shows the even order $r$ mode with $|m|=2, k=-2$ for $q = 10.01$. The vertical dashed line denotes the location of the critical angle $1/q$ ($\mu_{c1} = 1/1.33$ and $\mu_{c2} = 1/3.02$ on the top panel). Amplitudes of the Hough functions are set similarly to \cite{Lee_Saio_1997}.}
    \label{fig:extrahoughfunc}
\end{figure}

The top panel of Figure \ref{fig:hough and radial eigenfunctions} shows the absolute radial ($|\xi_r|$) and horizontal ($|\xi_h|$) displacement perturbations of the sixth radial order. Due to a relatively lower core temperature ($T_{\rm c} \approx 5$ million K), the WD core is about 8.4\% crystallized in radius, as indicated with the dashed line. The dependence of the full displacement eigenfunction, $\xi_r(r)\Theta(\theta)$, on both $r/R$ and $\theta$ in the
meridional plane is displayed in the bottom panel. Figure \ref{fig:extrahoughfunc} demonstrates the rotationally modified Hough functions associated with $|m| = 1, k = 0$ (retrograde $g$ mode) and $|m| =2, k=-2$ (retrograde $r$ mode) as functions of $\mu =\mathrm{\cos\theta}$. Dashed curves are for $q=1.33$, and the solid curves are for a relatively higher $q=3.02$ in the top panel. The $\mu=0$ or $\theta=90^\circ$ is being the equator and $\mu=1$ or $\theta=0^\circ$ being the pole of the star. The dashed lines indicate the respective $q$'s equatorial waveguide boundaries, which are the critical angles, 1/$q$. Most of the $g$-mode amplitudes are primarily confined below this angle and gradually disappear above the critical angle. For the $g$-mode, the degree of amplitude concentration increases towards the equator as the spin parameter increases. However, $r$-mode shows no such concentration. 
\begin{table}[ht]
    \centering
    \caption{Schema for mode identification}
    \begin{tabular}{c|c|c|c|cc}
    \hline \hline
    Angular & Azimuth & Prograde/ & $k$ & Mode type \\
    order& order & Retrograde  &  &  gravity ($g$)/&  \\
    ($\ell$) & ($m$) & &  & Rossby ($r$)&  \\ \hline \hline
    -& 1 & retro & -3 & $r$   \\ \hline
    -&  2 & retro & -2 & $r$  \\ \hline
    -&  1 & retro & -2 & $r$  \\ \hline
    -& 2 & retro & -1 & $r$ (Yanai)  \\ \hline
    -& 1 & retro & -1 & $r$ (Yanai) \\ \hline
    1& 1 & retro & 0 & $g$   \\ \hline
    1& -1 & pro & 0 & $g$ (Kelvin)   \\ \hline
    1& 0 & retro & 1 & Zonal   \\ \hline
    2 & 1 & retro & 1 & $g$  \\ \hline
    2& -1 & pro& 1 & $g$ (Yanai)   \\  \hline
    3& 1 & retro & 2 & $g$  \\ \hline
    3& -1 & pro & 2 & $g$   \\ \hline
    4& -1 & pro & 3 & $g$   \\ \hline
    \end{tabular}
    \label{table:table_mode_classification}
\end{table}

\subsection{Computation of Visibilities}\label{ssec:visibilitycomputation}
In order to compare the relative observability of modes, it is necessary to compute their relative visibility using their surface eigenfunctions and strength. Since we are not computing mode excitation in this work, we instead will compare the visibility of modes having equal kinetic energy (equipartition). In the comparisons below, the surface visibility of each mode depends on the limb-darkening coefficient ($\mu_l$) and the Eulerian pressure and temperature perturbations ($\delta P$ and $\delta T$), as given by \citep{Saio_2019}:
\begin{eqnarray}\label{eqn:visibilityform}
\nonumber
\text{Vis} \: &=& {\delta P\int_0^{2\pi} d\phi_L \int_0^{\frac{\pi}{2}} d\theta_L \, \sin(2\theta_L) \left[ 1 - \mu_l(1 - \cos \theta_L) \right]} {} \\
 && {\times \bar{\Theta}_{k}^{m} (\theta; q) \cos(m \phi)}
\end{eqnarray}
$\theta_L$ and $\phi_L$ are the spherical angles such that our line of sight corresponds to $\theta_L = 0$. $\Bar{\Theta}_{k}^{m} (\theta; q)$ is the normalized Hough eigenfunction such that $\int_{-1}^{1}[\bar{\Theta}_{k}^{m}(\mu)]^2d\mu=1$ and $\mathrm{sin(2\theta_L)}$ is the geometric factor. The kinetic energy of each mode is given by \citep{Aerts_2010}:
\begin{eqnarray}
\label{eqn:kineticenergy}
\text{K.E.} \:  &=& \frac{4\pi \omega^2}{2}\int_0^{R} \left[|\xi_{r}(r)|^2 + \lambda |\xi_{h}(r)|^2\right] \rho r^2 dr
\end{eqnarray}
The inclination, $i$, is the angle between the rotation axis and the line of sight. In other words, we need to normalize $\delta P$ by dividing by the kinetic energy of each mode to obtain the visibility strength.
\subsubsection{Surface Eigenfunctions}\label{sssec:surfaceeigenfunctions}

\begin{figure*}
  \includegraphics[width=.5\linewidth]{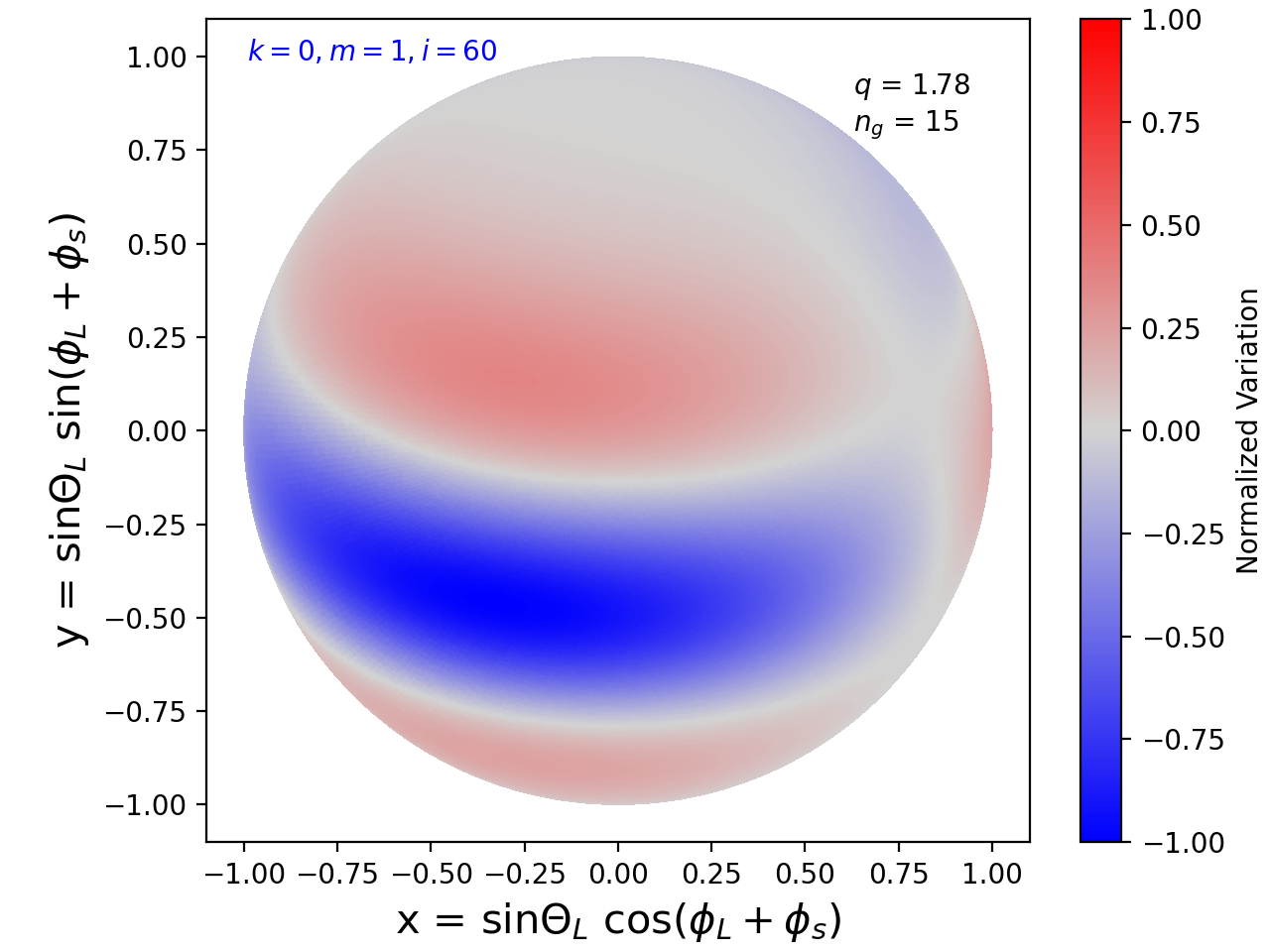}
  \includegraphics[width=.5\linewidth]{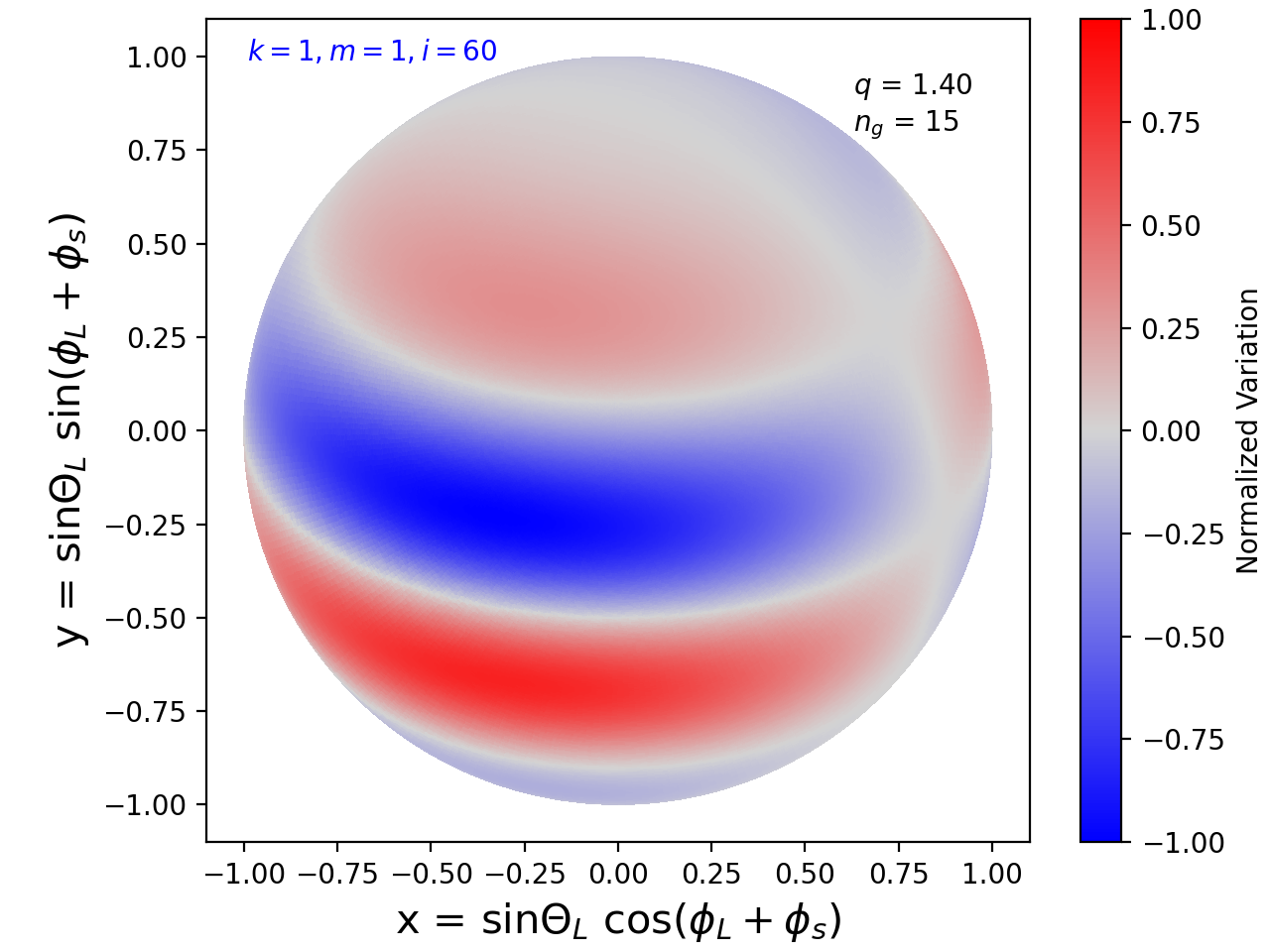}
  
  \includegraphics[width=.5\linewidth]{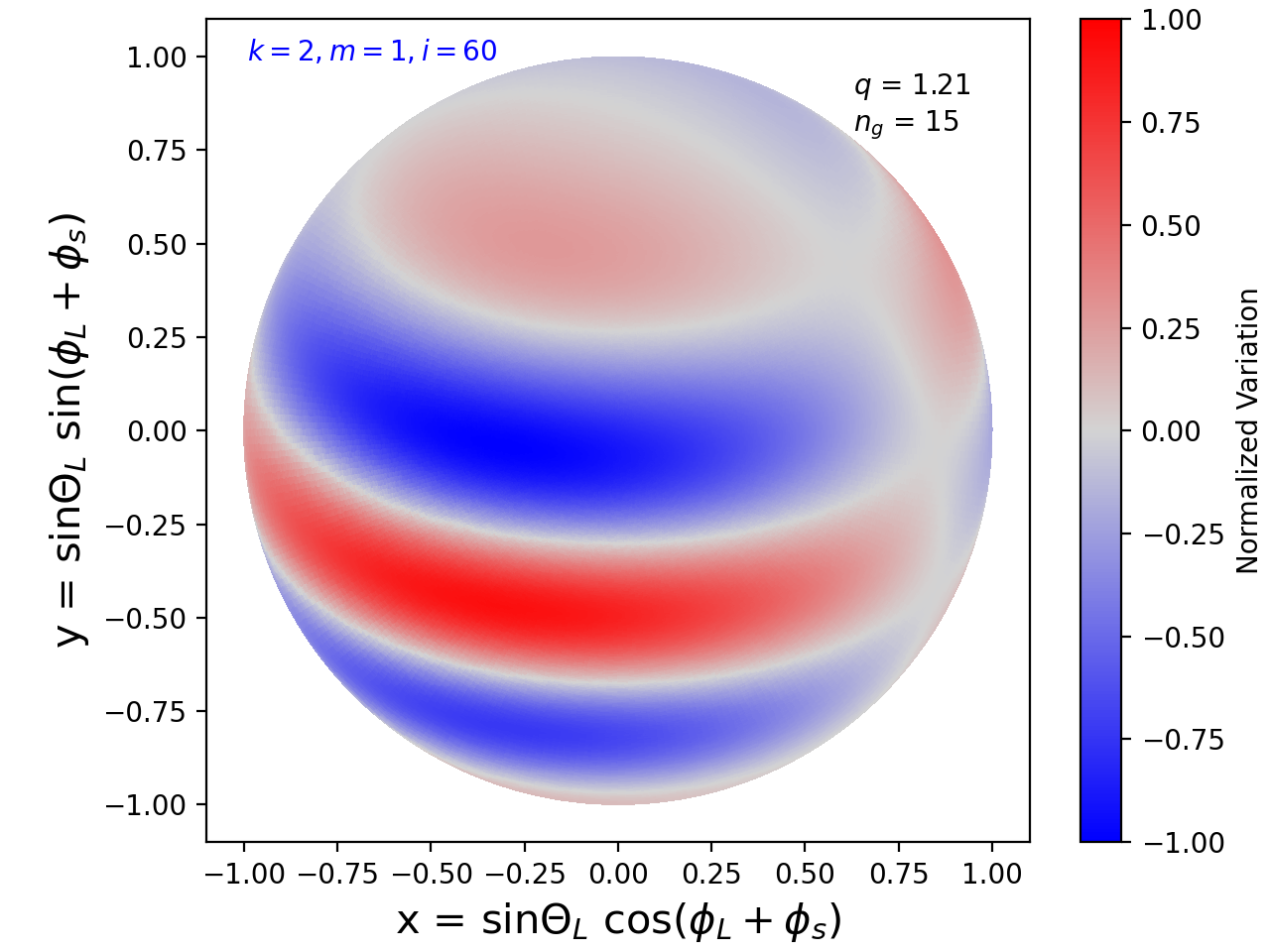}
  \includegraphics[width=.5\linewidth]{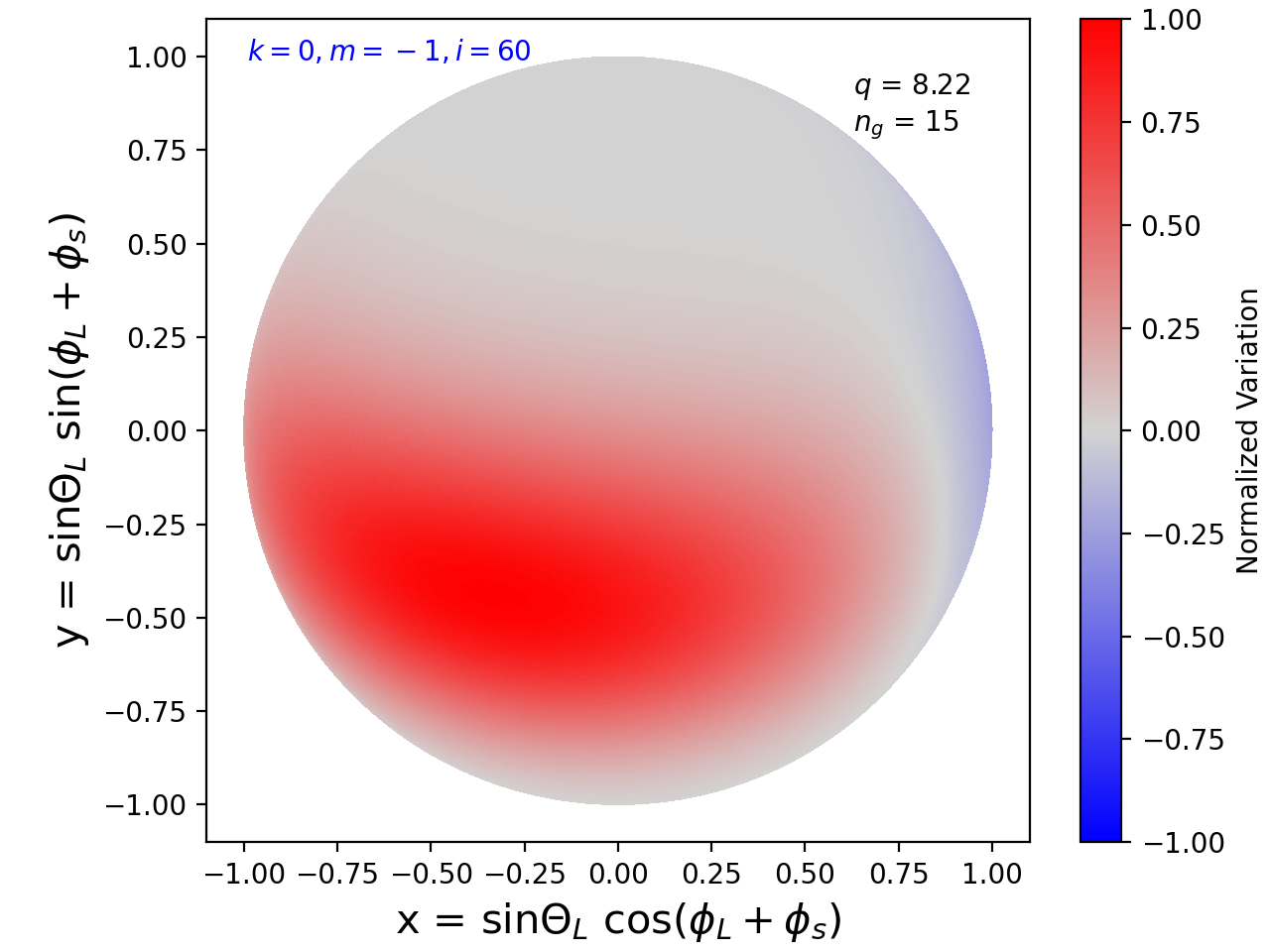}
  
  \includegraphics[width=.5\linewidth]{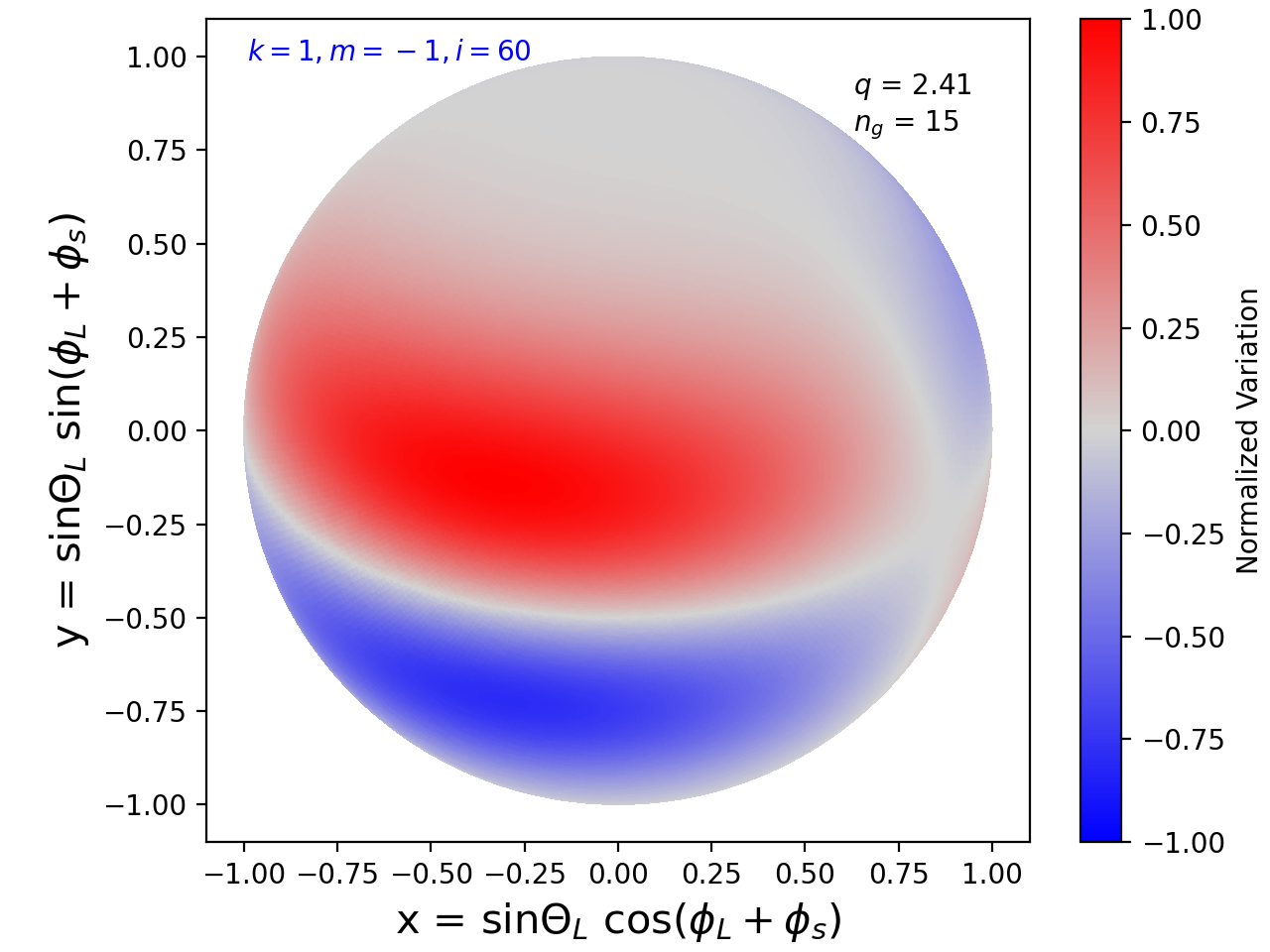}
  \includegraphics[width=.5\linewidth]{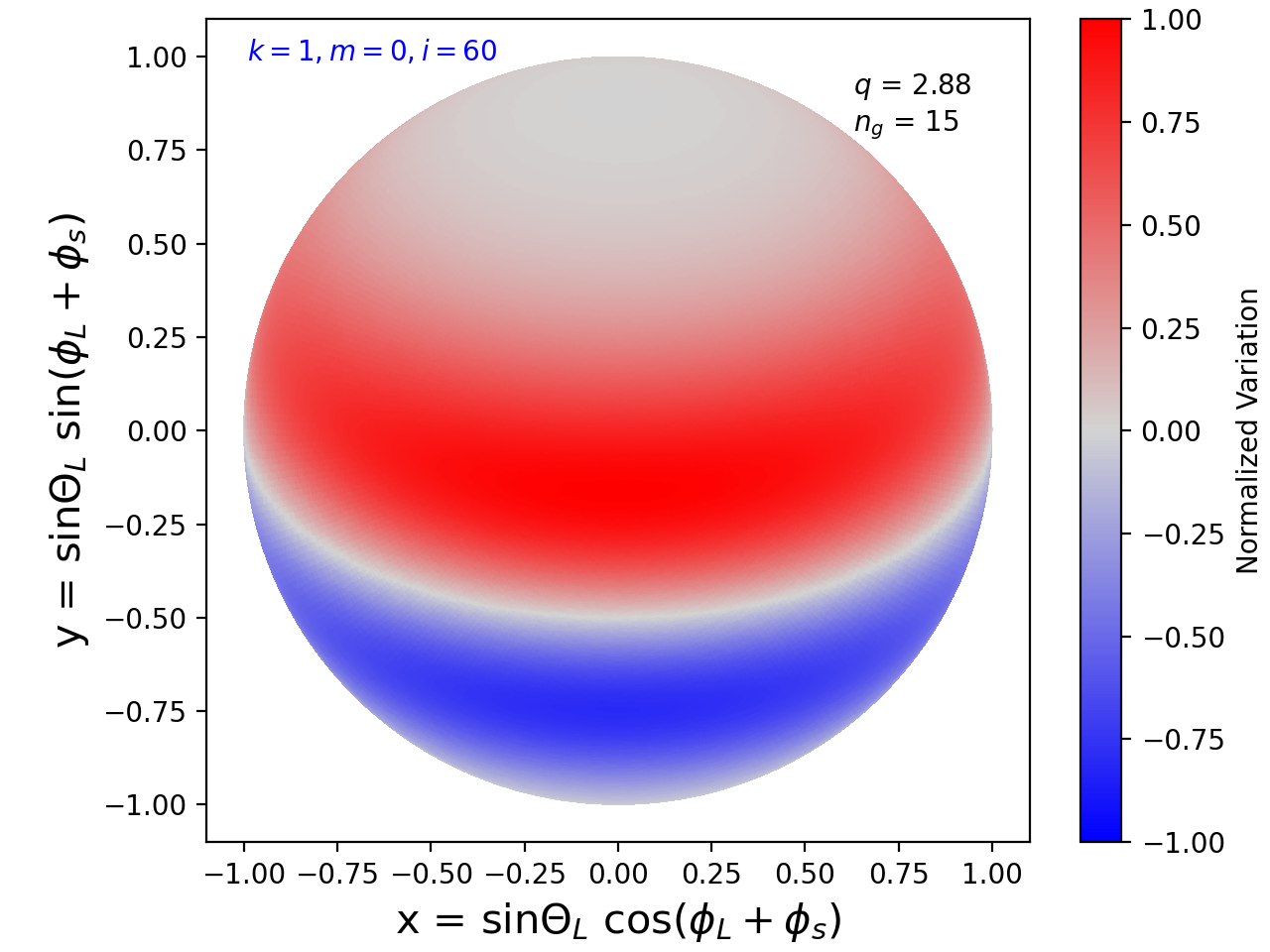}
   \caption{Distribution of pressure perturbations for a few selected $g$ modes, as labeled, on the star surface. The angle between the line of the sight and the rotation axis (the inclination) is set to be 60$^\circ$ for all cases. The x and y axes are projected linear offsets from the line of sight, with $\phi_s=\pi/2$ chosen so that the rotation axis appears pointing upward. Positive, negative, and zero variations are indicated with red, blue, and gray, respectively. The spin parameter ($q = 2\Omega/\omega$) and $k$-values are labeled on each plot. The visibility for $g$ mode peaks at/around the radial order $n_g = 15$ for all the inclinations displayed (see figure \ref{fig:gmodevisibility}).}
    \label{fig:surfacegmode} 
\end{figure*}
\begin{figure*}[ht]
  \includegraphics[width=.5\linewidth]{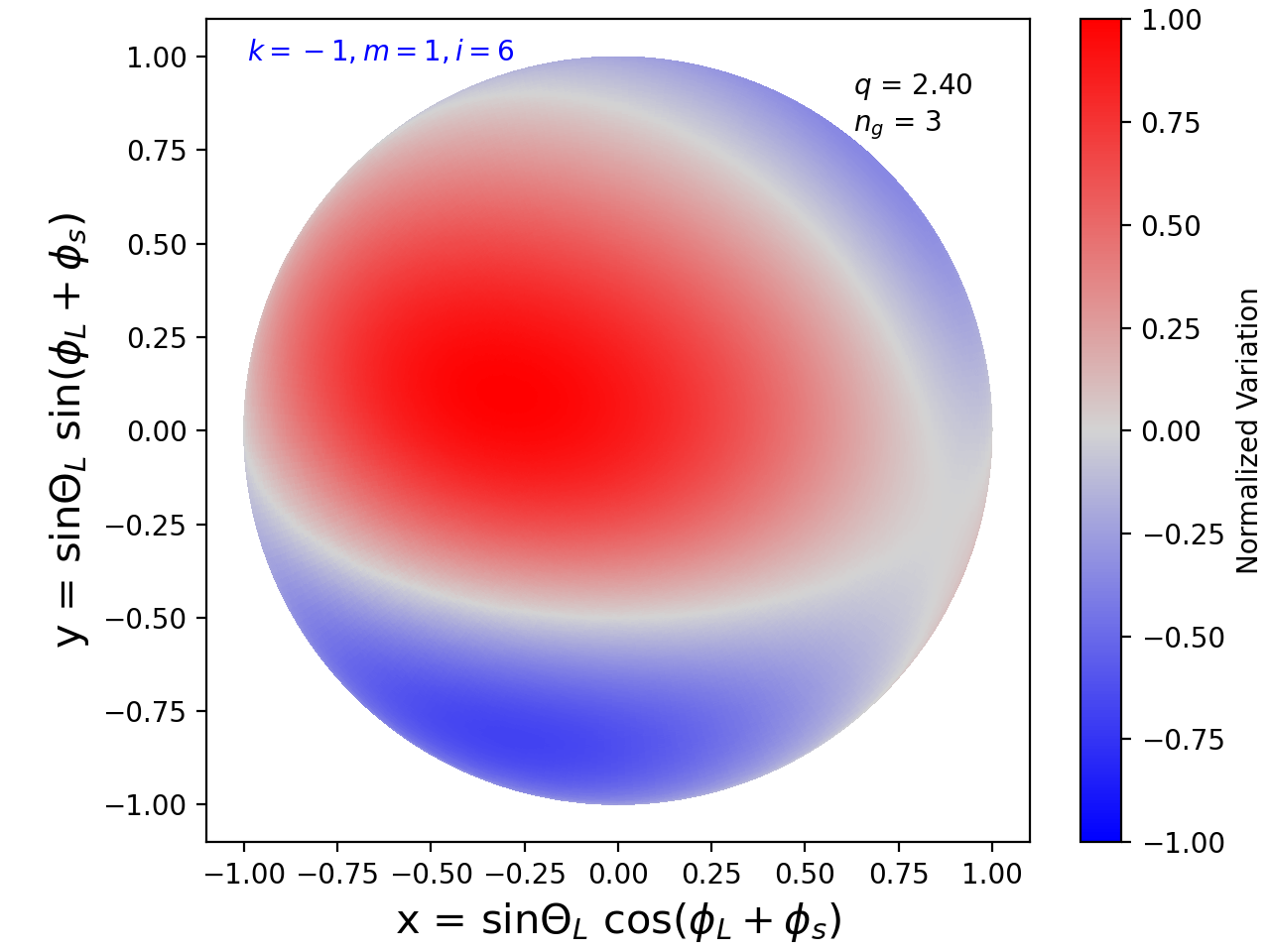}
  \includegraphics[width=.5\linewidth]{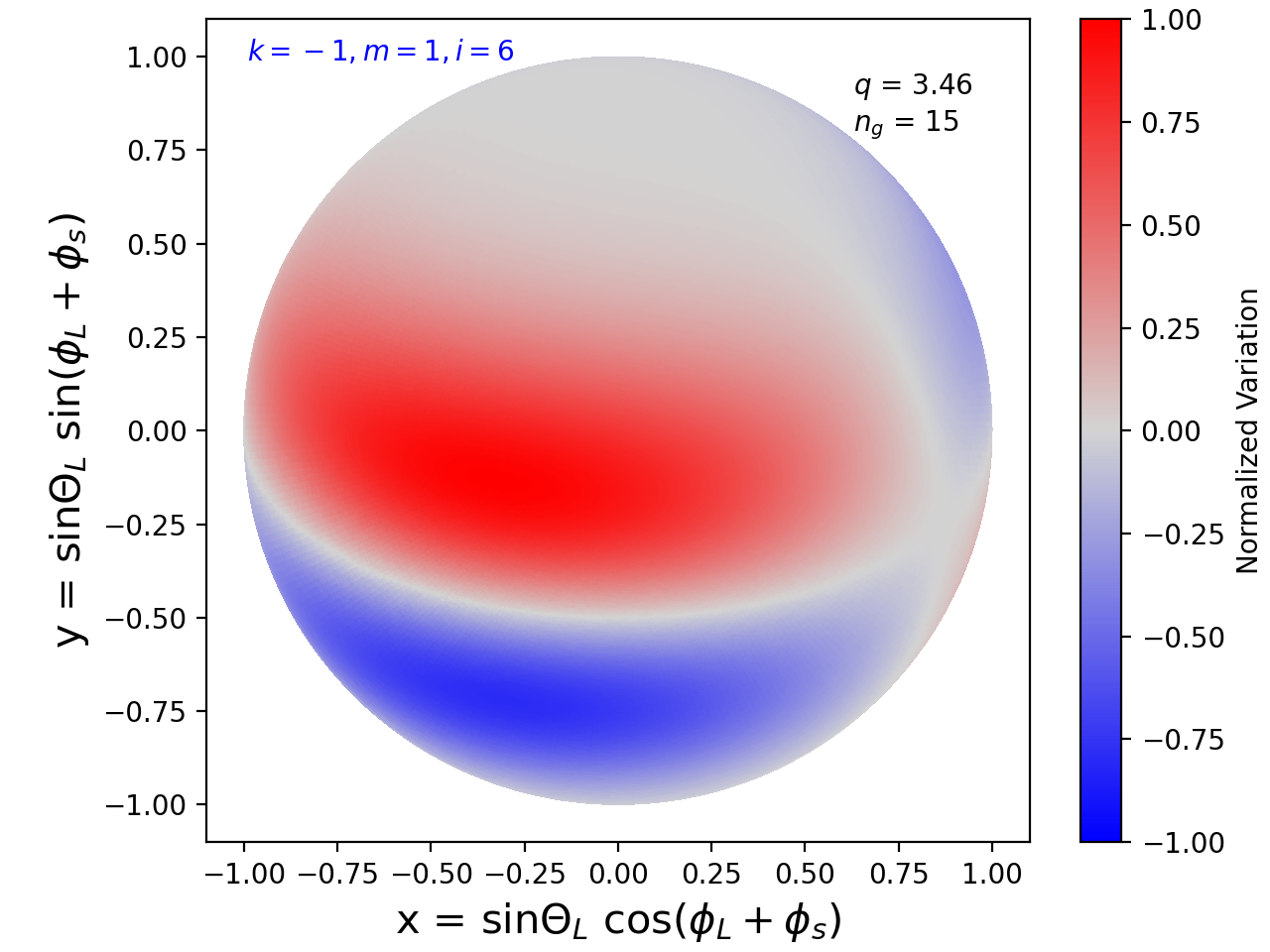}
  \includegraphics[width=.5\linewidth]{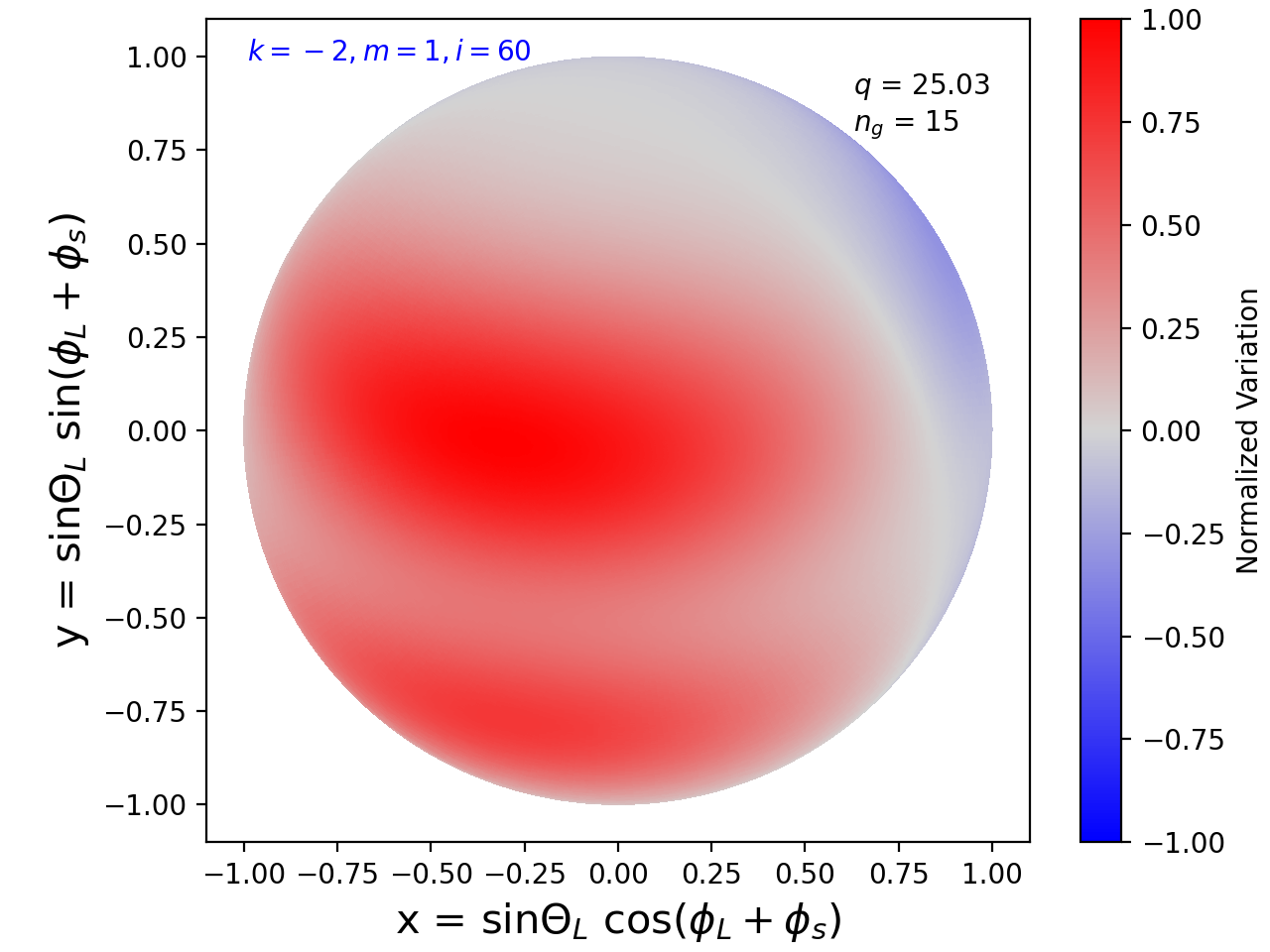}
  \includegraphics[width=.5\linewidth]{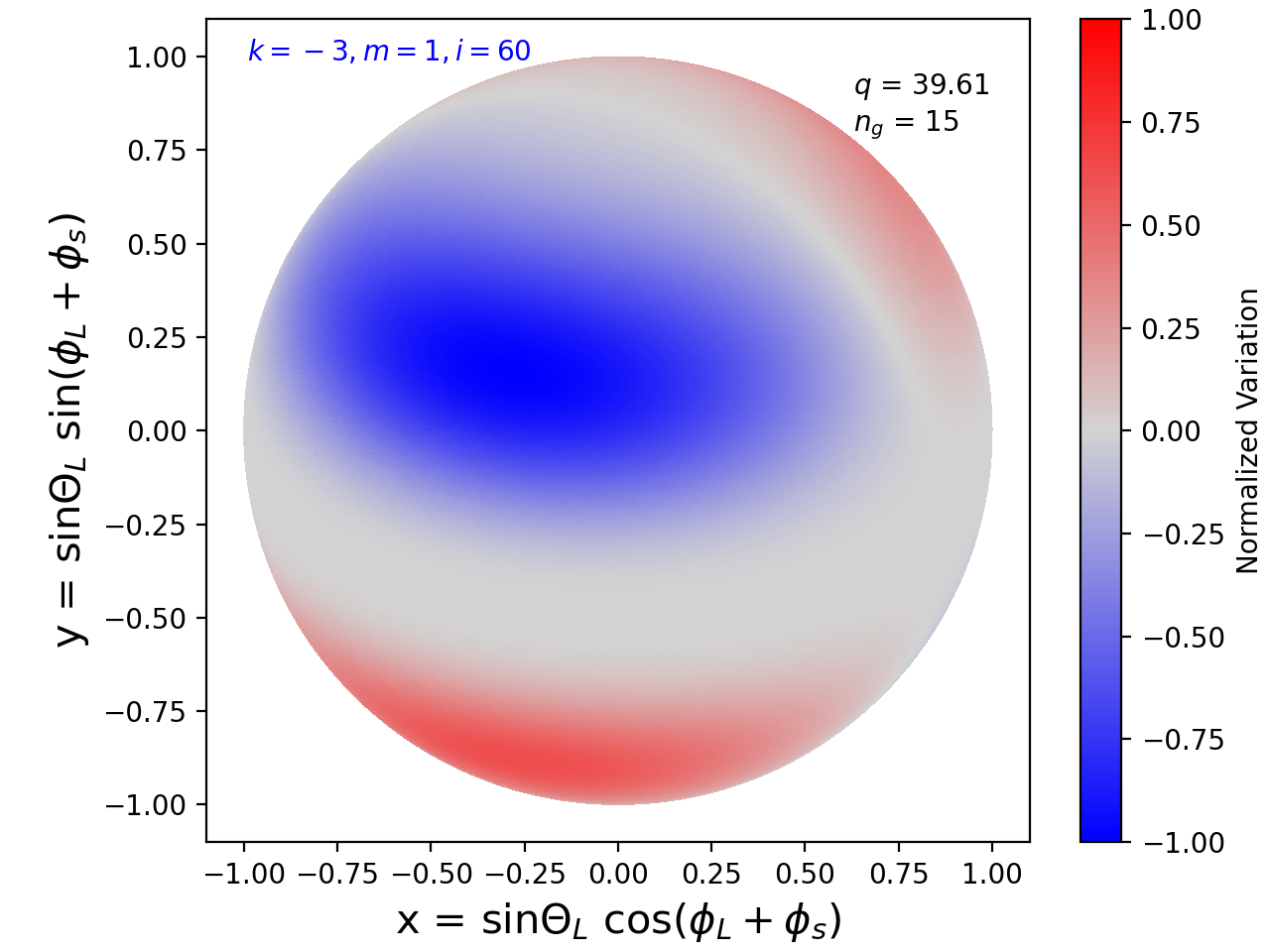}
  
  \includegraphics[width=.5\linewidth]{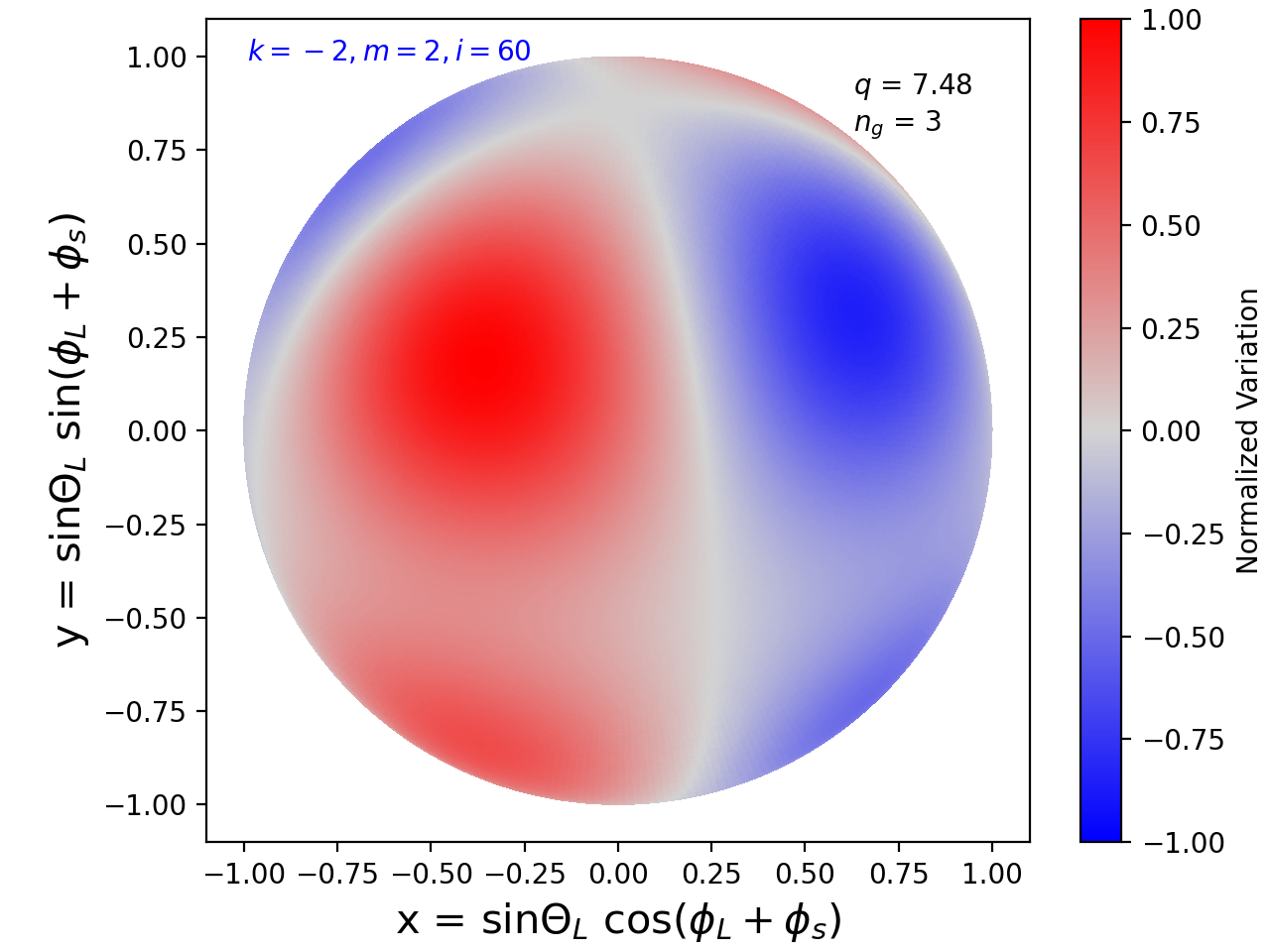}
  \includegraphics[width=.5\linewidth]{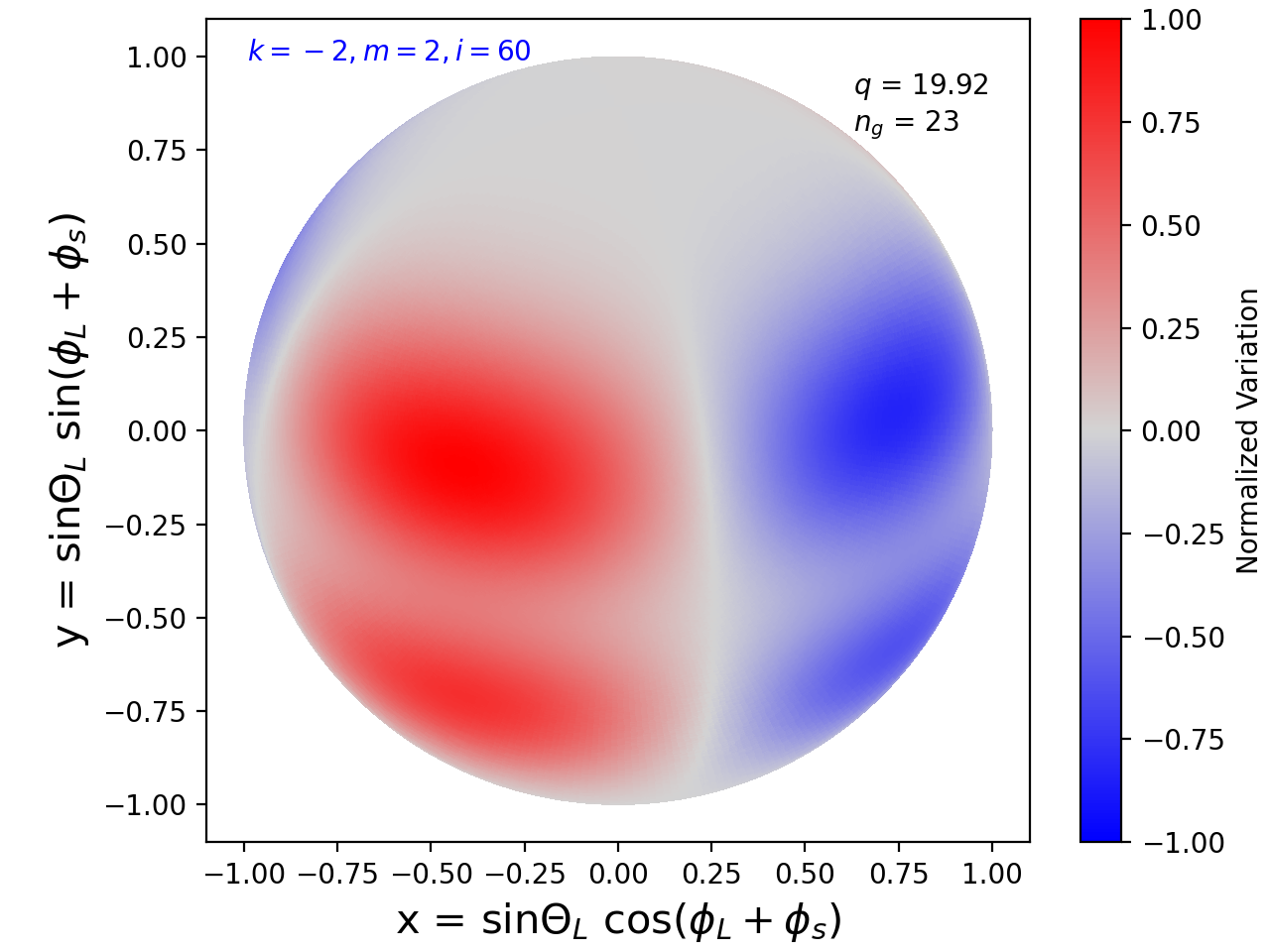}
   \caption{Similar to figure \ref{fig:surfacegmode} but for Rossby modes. The odd and even $k$ values represent the odd and even order Rossby modes, respectively.}
   \label{fig:surfacermode}
\end{figure*}
The distributions of pressure perturbations on the stellar surface for certain $g$ and $r$ modes with an inclination of 60$^\circ$ are shown in Figures \ref{fig:surfacegmode} and \ref{fig:surfacermode}. Red (blue) indicates the positive (negative) perturbation variations. The gray area represents the zero perturbation variations (nodal planes). Each plot is labeled with the respective $q$ value, ordering index $k$, azimuthal order $m$, and radial order $n$. The modes shown are drawn from the actual mode set of the stellar structure introduced in section \ref{sec:wdmodel}, also used in the previous and following subsections. The variations are normalized to their maximum. At larger spin parameters, the amplitude is typically focused toward the equator. 

A substantial correlation exists between the spin parameter and $g$ mode surface eigenfunctions. The surface amplitude differs considerably in retrograde and prograde modes at similar spin parameters for the same mode order. Figure \ref{fig:surfacegmode} shows that for the lowest azimuthal order $k=1$, the retrograde mode (top right) has an extra node on the star surface compared to the prograde mode (bottom left). This is well supported by Figure \ref{fig:lambdaspin}, with retrograde modes having about an order of magnitude larger value of $\lambda$ than the prograde modes. This suggests retrograde $g$ modes will have more cancellation of surface brightness variations and thus be more difficult to observe. Figure \ref{fig:surfacegmode} also indicates that $g$ mode amplitude is easily modified under the influence of rotation and is strongly squeezed towards the equator at larger $q$. 

As seen in figure \ref{fig:surfacermode}, the $r$ mode order $k=-1$, $|m|=1$ visibility amplitude depends on the spin parameter. The amplitude is modestly concentrated towards the equator if $q$ is much higher than 2. This feature is supported by Figure \ref{fig:lambdaspin}, as at higher $q$'s, the $\lambda$ becomes very large. However, the even-order modes, with $k=-2$, $|m|=1$, and $k=-2$, $|m|=2$, are less strongly affected by the spin parameter, requiring a much higher spin parameter in order to shift amplitude away from the pole. Figure \ref{fig:surfacermode} demonstrates that the amplitude of these even-order modes is constrained towards the mid-latitude region of the star. Additionally, this is supported by eigenfrequencies of LTE becoming nearly constant at higher spin parameters. Kepler observations suggest that these stars should be examined at moderately high inclination angles, where mode visibility and amplitude modulation become more pronounced \citep{Van_Reeth_2016}.

\subsubsection{Visibilities for all the modes vs. frequency}\label{sssec:visibilitiesofallmodes}

\begin{figure*}[ht]
    \includegraphics[width = 1\linewidth]{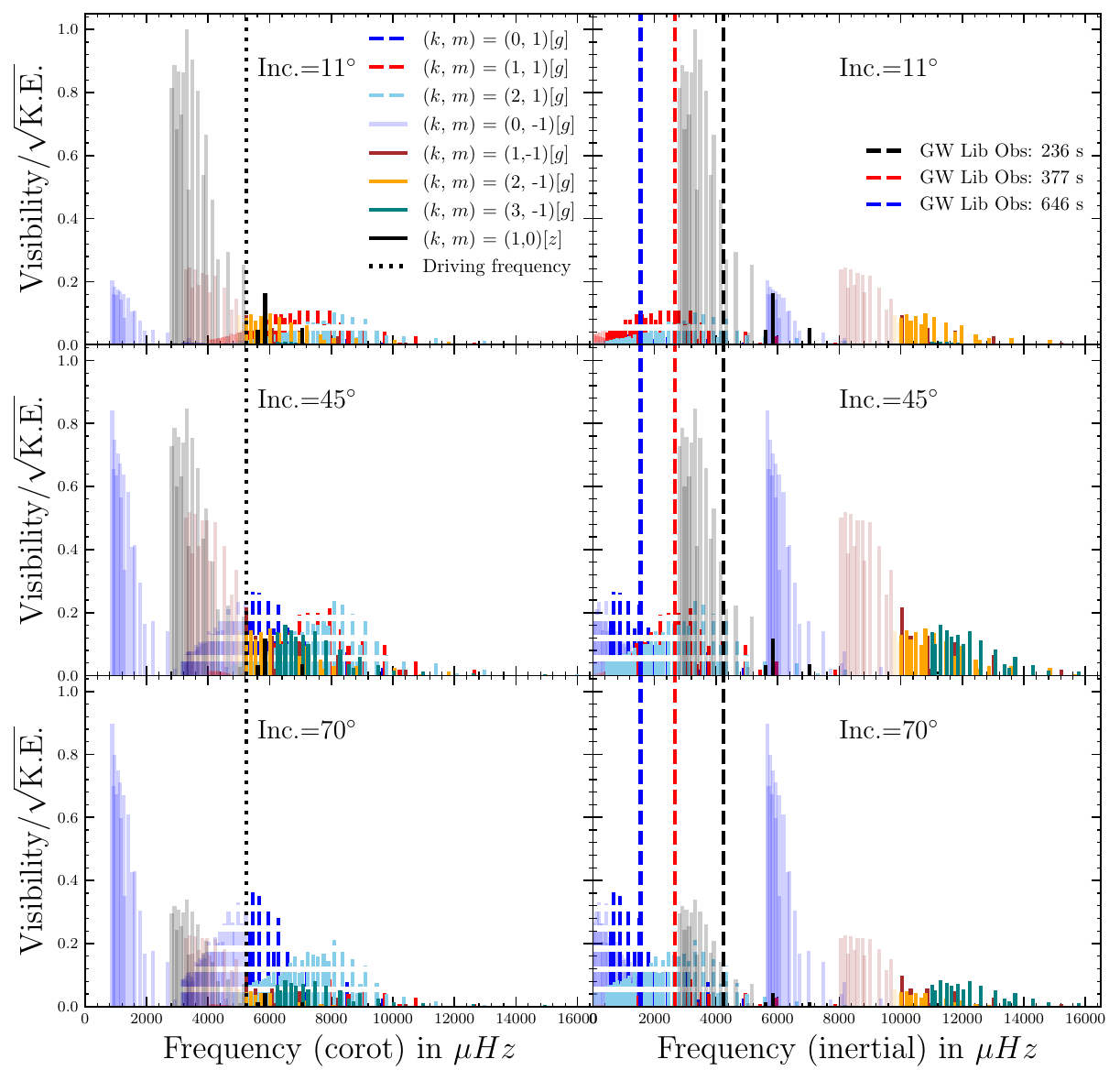}  
    \caption{Expected normalized visibility of gravity modes in both the star's frame (left panel) and the observer's frame (right panel) at different inclinations. Each mode order is scaled by kinetic energy in order to mimic what might be seen if equipartition were realized. The visibilities have been divided by the maximum value for both $g$ and $r$ modes for a fixed viewing angle. Dashed and solid lines are retrograde ($m > 0$) and prograde ($m < 0$), except for the solid black lines corresponding to zonal modes ($m = 0$). The vertical dotted line at approximately 5200 $\mu$Hz (period = 191 s) indicates our hypothetical driving scenario in which modes with frequencies higher than this threshold in the corotating frame are driven. These modes are drawn with stronger shading in both panels. The Observed GW Lib frequencies are marked with the vertical dashed lines (right panel).}
    \label{fig:gmodevisibility} 
\end{figure*}

\begin{figure*}[ht]
    \includegraphics[width=1.\linewidth]
    {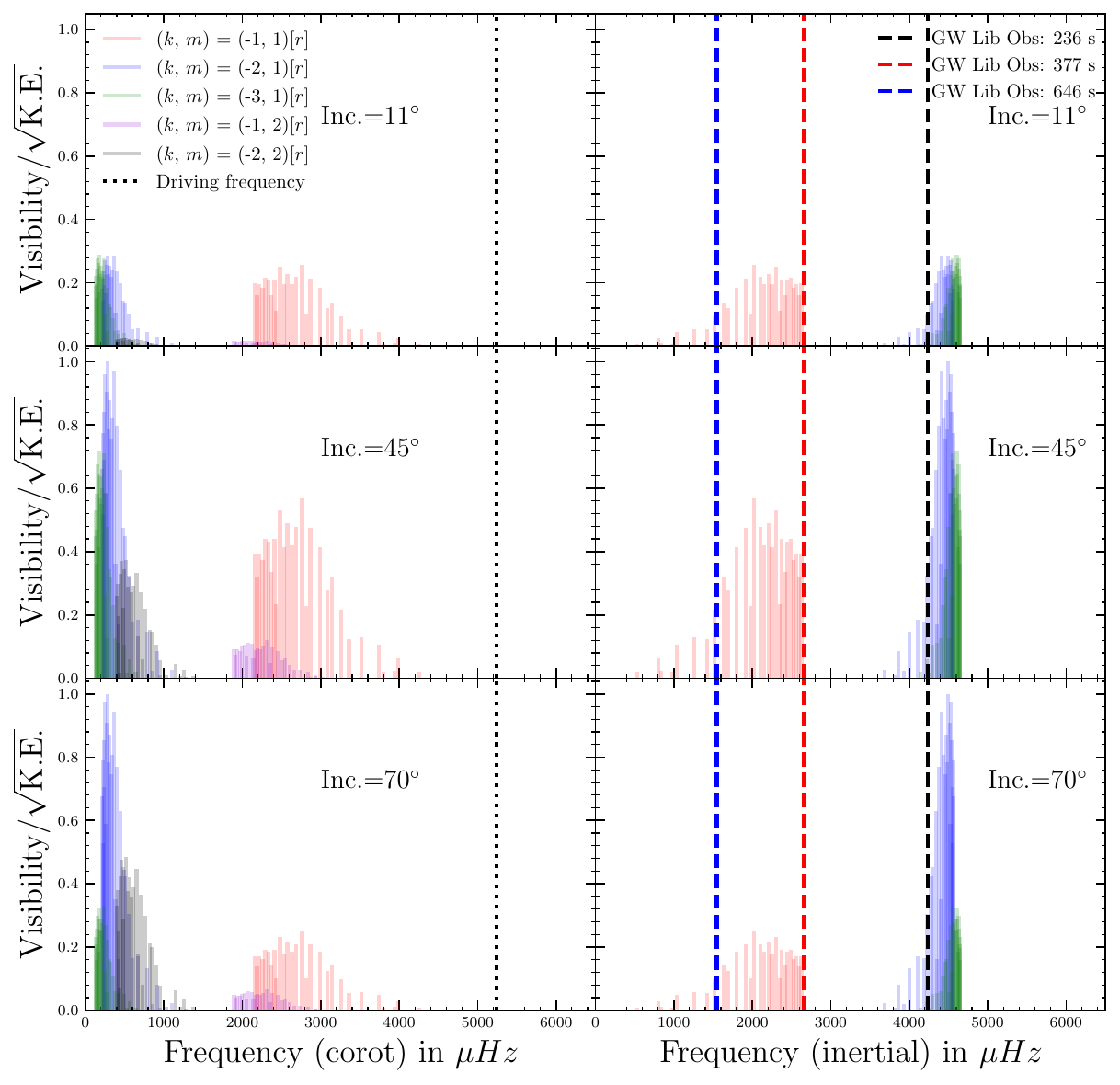} 
    \caption{Same as figure \ref{fig:gmodevisibility} but for retrograde Rossby modes. The overall visibility is normalized to the maximum value of $g$ and $r$ modes. Note a significantly smaller range of frequencies is shown here compared to Figure \ref{fig:gmodevisibility}.}
    \label{fig:rmodevisibility} 
\end{figure*}
In order to be able to compare mode visibilities in the absence of a computation of driving and damping, we approximate an effective equipartition. This is accomplished by computing the visibility (eq. \ref{eqn:visibilityform}) and kinetic energy (eq. \ref{eqn:kineticenergy}) for each mode with an arbitrary normalization, then dividing the visibility by the square root of the kinetic energy. For each inclination, both $g$ and $r$ modes are normalized separately. This approach is similar to the comparison made in \citep{Saio_2019}. Once this is done for all modes, the resulting $\text Vis./\sqrt{\text K.E.}$ is then again divided by the highest value obtained for all modes shown in Figures \ref{fig:gmodevisibility} and \ref{fig:rmodevisibility}, for each inclination separately. In the present case, the maximum visibility corresponds to the $n_g=20$, $k=-2$, $m=1$ $r$ mode at viewing angle $i=45$. This allows us to compare all modes under the equipartition assumption without having to posit a relationship between the surface pressure perturbation $\delta P$ and the resulting surface brightness variation. 

Figures \ref{fig:gmodevisibility} and \ref{fig:rmodevisibility} display the predicted surface visibilities, using equation \ref{eqn:visibilityform}, over the square root of kinetic energy of various mode orders for $g$ and $r$ modes, three months after the outburst, in the star's frame (left panels) and the observer's frame of reference (right panels), respectively. Recall that the inertial frame frequencies, $\omega_i$, observed by a distant observer via brightness variation from a single line of sight, are related to the frequencies in the star's co-rotating frame, $\omega$, by $\omega_i = \omega - m\Omega$. The black vertical dashed line and the line weight of all the individual lines are related to the mode driving scenario and discussed in section \ref{ssec:observablesignificance}.

In figure \ref{fig:gmodevisibility}, prograde and retrograde $g$ modes are shown by solid and dashed lines, respectively. For the $g$ modes, the highest visibility amplitude mode is between the radial orders 10 and 20, depending on the viewing angle. For the $r$ mode, the amplitude peaks stay between the radial orders 15 and 20. The zonal mode, $k=1,m=0$ (solid black line), shows the most net surface variation when viewed from the pole (see bottom right panel in Figure \ref{fig:surfacegmode}) and thus is most visible at a smaller inclination (pole at $0^\circ$), i.e., highest visibility at $11^\circ$ in Figure \ref{fig:gmodevisibility}, with decreasing visibility at higher inclination due to cancellation
when viewed from equatorial lines of sight. At $45^\circ$, the prograde mode, $k=0,m=-1$ (solid blue), is the most visible mode (excluding the zonal mode) compared to $k=1,2$. This even $k$ mode has the highest visibility for equatorial lines of sight, since there is less cancellation (see middle right panel of Figure \ref{fig:surfacegmode}), and lower visibility for polar lines of sight. This is a general feature of even-order modes, making them prominent at equatorial lines of sight. The 3rd highest visibility modes at 45$^{\circ}$ inclination are the $k=1$, $m=-1$ modes (solid brown). Cancellation leads these odd $k$ modes to have lower visibility at both higher and lower inclination (see bottom left panel in Figure \ref{fig:surfacegmode}). The general trend continues that even $k$ modes are more prominent at equatorial lines of sight and odd $k$ modes are most prominent at mid-inclination lines of sight. However, for modes beyond the lowest order, the relative prominence depends on the details of the individual surface eigenfunctions. Specifically, the additional node planes present in the retrograde modes compared to prograde modes with the same $k$ (compare top right and bottom left panels in Figure \ref{fig:surfacegmode}, both for $k=1$) lead to differences in their visibilities at different inclinations. Retrograde modes generally show more cancellation at a given $k$ (compare the red dashed and brown solid lines in Figure \ref{fig:gmodevisibility}). 

For non-polar lines of sight, the predicted visibilities of the strongest $r$ modes generally exceeds than the strongest $g$ modes. For instance, at an inclination of 45$^\circ$, the peak visibility of the lowest-order even $r$ mode is approximately 20\% higher than that of the dipole $g$ mode. Recall that, for each inclination shown in figures \ref{fig:gmodevisibility} and \ref{fig:rmodevisibility}, the visibilities are scaled by a single overall factor so that the highest visibility is unity for the respective inclination angle. Figure \ref{fig:rmodevisibility} shows that for $m= 1$ $r$ modes, even $k=-2$ has higher visibilities than the odd $k=-1,-3$. Compared to the odd-order $r$ modes, the even $r$ modes have a greater visibility amplitude as the inclination increases. Remarkably, the $k=-2$, $m=2$ mode shows higher visibility at an inclination of $70^\circ$ than the lowest order $r$ mode, which is odd order. This closely resembles the visibility computation made by \cite{Saio_2019}.

Figures \ref{fig:gmodevisibility} and \ref{fig:rmodevisibility} also illustrate that some modes of consecutive radial order within a given $k$, $m$ series exhibit larger visibility than others at similar frequencies. This is because, for a fixed surface amplitude, some modes have a larger amplitude in the core. These modes have a larger kinetic energy, so that when we divide the integrated surface perturbation by the kinetic energy, in order to represent approximate equipartition, higher K.E. modes end up having lower visibilities. This same phenomenon has been discussed for isolated WDs, where it is generally referred to as mode trapping. Modes with relatively small core amplitudes are said to be trapped in the envelope \citep{Brassard_1992, Charpinet_2000}. We will not undertake a classification of members of the mode sequence, just remarking that this is the source of the observed variation in visibility along the sequence. This work calculates visibility three months after the dwarf novae, shortly after the WD quiescence phase. Since, as shown in \cite{Kumar_Townsley_2023}, the eigenfunctions shift modestly as the surface layer cools, the relative visibilities of modes of similar radial order may also change. This may offer an interesting window into the mode character but may be difficult to observe due to the complexity of mode driving, i.e., modes may not be close enough to strict equipartition.

\subsection{Observable Significance}\label{ssec:observablesignificance}
Figures~\ref{fig:gmodevisibility} and \ref{fig:rmodevisibility} indicate that at higher inclination (lines of sight nearer the equator), the lowest-order even $r$ mode ($k=-2$, $m = 1$) exhibits a stronger visibility amplitude than the even $g$ mode with $k=0, m = -1$. Period spacings and period patterns of $\gamma$ Dor stars reveal that most Rossby modes are found to be $k=-2$, $m=1$ mode \citep{Li_2019}. This is consistent with our visibility calculations, despite $\gamma$ Dor stars having very different stellar structures from the stars studied here. This is a property specific to the surface eigenfunctions and validates our visibility calculations. This contrast of having higher visibility for $r$-modes, for these two modes, and generally, is consistent with the general thesis put forward by \cite{Saio_2019}: that $r$-modes should be more prominent than $g$-modes. However, by explicitly computing the visibilities of both mode families, we do not find the contrast between $r$ and $g$ mode visibilities to be strong enough that $g$ modes should be altogether unobservable. There is also an issue of driving, which we address next. 

Gravity modes in isolated white dwarfs are primarily driven by convection through ``convective driving" \citep{Brickhill_1991, Wu_Goldreich_1998}. Energy is pumped into the mode through the interaction of its periodic compression of the surface with the surface convection zone related to the ionization of H or He. 

For accreting WDs with low mass star companions, the material at the surface is hydrogen rich, having recently arrived from the companion. While more detailed analysis is available \citep{Arras_Townsley_Bildsten_2006, Van_Grootel_2015}, we make a simple comparison to the pure H case. Due to the presence of He, this convection zone is the appropriate depth for driving the observed frequencies when the star is slighty at higher $T_{\rm eff}$ ($\sim$ 14,000 K) than that appropriate for pure H ($\sim$ 11,000 K) \citep{Arras_Townsley_Bildsten_2006}. The thermal timescale at the bottom of the convection zone is the important metric to understand driving. 

Although we will not undertake a direct analysis of convective driving in this work, it is useful to consider a driving threshold of similar character to what might arise from convective driving. In the convective driving framework, modes of shorter periods (higher frequencies) than the thermal time at the base of the convection zone are driven. In order to consider such a situation qualitatively, we will choose a representative threshold that is in the middle of the range of our computed modes. This is intended to represent a scenario in which a modest number of modes are driven. In this case we choose a frequency of 5234 $\mu$Hz (period of 191 seconds). Modes with higher frequencies in the star's frame than this will be taken as being driven. This threshold frequency is indicated by the vertical line in Figures \ref{fig:gmodevisibility} and \ref{fig:rmodevisibility}, and the lines for modes below this threshold are shown lighter than the driven modes above this threshold. Note that these modes shift to different frequencies when observed depending on whether the mode is prograde or retrograde. Figure \ref{fig:rmodevisibility} illustrates that the $r$ modes are less likely to be observed because, even though they are shifted to frequencies like those observed and have higher visibility than the $g$ modes, they are unlikely to be driven when there are only a handful of modes visible because they would be well below the minimum driven frequency.

The three observed GW Lib mode periods, 236 s, 377 s, and 646 s, are marked with the three vertical dashed lines in the observer's frame in figures \ref{fig:gmodevisibility} and \ref{fig:rmodevisibility} \citep{Thorstensen_2002}. While these periods seem to appear near the zonal mode regions in the visibility plot and coincide with the strong visibility on the star's surface (Fig. \ref{fig:gmodevisibility}, right panels), they are unlikely to be given enough energy to drive them in the observer's frame due to their lower mode frequencies. Instead, they closely resemble $k=1$ and $k=2$ retrograde modes with $m=1$ at higher inclinations. Even though the retrograde modes possess more cancellation on the star's surface, as illustrated in Figure \ref{fig:surfacegmode} than their prograde counterparts, as the retrograde mode is having an extra node on the star's surface for the similar mode order and spin, we believe that the observed mode periods of GW Lib may correspond to retrograde $g$-modes. It also may be the case that the highest frequency visible mode in GW Lib corresponds to one of the lowest radial order zonal modes, while the two lower frequency modes correspond to the retrograde mode groups. This may provide an explanation for why some modes appear more stable in frequency than others, since closely spaced mode groups with time-varying amplitudes may appear as less stable modes. Although we emphasize that we are not explicitly fitting or matching the GW Lib observed mode periods within our calculation, rather we align the mode periods based on their relevance to surface visibility within the known mode driving scenario. Achieving a proper mode fitting requires further investigation, including a more comprehensive model and a broader set of observed modes.

\section{Validity of Traditional Approximation of Rotation} \label{sec:tarvalidation}
Computation of full eigenmodes with the impact of stellar rotation is a well-known challenge. For rotating stars, the pulsation equations system is inseparable in the radial and latitudinal coordinates ($r$, $\theta$). TAR greatly simplifies the understanding of Coriolis force effects restoring the separability of the problem, especially on low-frequency high order $g$ modes \citep{Townsend_2003a}. In particular, it assumes a solid body rotation. Due to uncertainties in the angular momentum transport mechanism, differential rotation is not explored and would significantly increase the computational complexity. Without rotation, the angular eigenfunctions of LTE are the spherical harmonics; with rotation, they are the Hough functions \citep{Lee_Saio_1987}. This approximation is reasonable for slow to moderate rotators with spin parameter $q\le 2$ \citep{Ballot_2012}, including for the computation of $r$-modes. Furthermore, this approximation fails to determine the mode character near the resonance where gravito-inertial and pure inertial mixed mode characters are present \citep{Ouazzani_2020}. In that work it was found that although TAR reproduces arguably complete calculations for modes confined in the radiative region of the star, it completely fails to treat mode character related to the convective region. 

For the rapidly rotating accreting systems such as GW Lib, EQ Lyn, and more, the TAR framework is inadequate to model the pulsation modes correctly near the central part of the star. This is due to the fall-off of buoyancy near the star’s center. A large enough solid core can exclude modes from this region, but in developing models for this work we have found that much of the interesting parameter space is likely to not have a sufficiently large solid core. Even further, seismology might be an important way to confirm whether or not the accreting WDs have a solid core. That would necessarily require accurate computation in cases that lack a solid core.

The top panel of figure \ref{fig:hough and radial eigenfunctions} shows the eigenfunctions of the radial ($\xi_r$) and horizontal ($\xi_h$) displacements of a gravity mode of a 0.78~$M_\odot$ WD model, of sixth radial order. The ratio, $\xi_r$/$\xi_h$, is small, as TAR requires, in the outer part of the star but large, even $> 1$, near the core (note the important amplitude is that in regions away from the nodes). Thus the TAR fails in this region even with a solid core. Although the anticipation is that, even so, the global mode behavior may show similar features qualitatively \citep{Kumar_Townsley_2023}, quantifying frequencies requires the computation of the non-approximate mixed gravito-inertial modes. However, since the TAR applies to a large fraction of the star, it is
expected that essentially all $g$ and $r$ modes found here will each have a corresponding mode in the inseparated, non-approximate eigenmode solutions. Thus, this work is expected to be useful in identifying and classifying modes both while constructing solutions to the coupled eigenmode problem in $r$ and $\theta$ and while considering the observed modes until those non-approximate solutions become available. 

\section{Conclusions and Future Work} \label{sec: discussionandconclusions}
For the first time, we have simultaneously investigated $g$ and $r$ modes in the rapidly rotating accreting WDs, performing the full visibility computations that account for the distribution of surface variations. 
Our main results are summarized as follows:
\begin{itemize}
    \item Using the eigenmode visibility calculations within the framework of TAR and the expected driving mechanism \citep{Arras_Townsley_Bildsten_2006, Van_Grootel_2015}, we find that $g$ modes are more relevant to observations. Our $r$-mode visibility results are consistent with \cite{Saio_2019}, among the low-order modes, the even-order $k=-2$ shows the strongest visibility without the consideration of mode driving. However, we find that the several $g$-modes achieve comparable visibilities, with peak amplitudes typically 20-60\% of the maximum, and are not significantly different than the less dominant $r$-modes. 
    Efficient mode excitation requires mode periods in the corotating frame to be comparable or or shorter than the convective turnover timescale at the base of the convective envelope of the star ($\tau_{\rm cvz}$). In the accreting WDs, as the surface heats due to accretion, $\tau_{\rm cvz}$ closely matches the periods of $g$-modes, which are shorter than the most visible $r$-modes. This suggests that the observed pulsations in accreting WDs are most likely gravity modes, unless an alternative effective excitation mechanism is capable of driving $r$-modes.

    \item Additionally, we compare the visibility of $g$ and $r$ modes at different inclinations, considering both pole-on and equator-on orientations. Gravity modes show a strong sensitivity to stellar rotation, which confines the amplitudes towards the equator. As a result, even-order $g$ modes exhibit stronger visibilities when observed near the equator, while odd-order modes are affected by the cancellations due to the antisymmetric nature of their surface eigenfunctions. In contrast, $r$ modes are comparatively less affected by stellar rotation. As illustrated in Figure \ref{fig:lambdaspin}, the $\lambda$ values for higher order modes (with the exception of $k=-1$ mode) remain nearly constant. 
    

    \item Neither of $g$ nor $r$-modes have a clear advantage in visibility, low-order high frequency modes $g$-modes are likely to be observed. The dominant set depends on the inclination, and the tight spacing of retrograde $g$ modes may explain the apparent lack of stability seen in some periodicities in GW Lib. A full mixed-mode treatment would replace both the (approximated) $g$ and $r$ modes with a unified set of modes, with modes showing varying $g$- and $r$-like character and visibility. We plan to explore this in our future work.

    \item We adopted a realistic accretion rate and included element diffusion during the long-term accretion phase, significantly improving our accreting WD model over \citep{Kumar_Townsley_2023}. Despite these structural refinements, we find similar mode spacing properties to those reported in \citep{Kumar_Townsley_2023}, with alternating small-large period spacings as a robust feature of the mode spectrum. Additionally, we assess the applicability and feasibility of TAR in the rapidly rotating accreting WD with a solid core.  Although, the presence of a sufficiently large solid core can exclude the modes from that the central vicinity, we find that not all cases that we expect need to be considered in a seismological fit will have a sufficiently large solid core.
\end{itemize}
\acknowledgments
This manuscript has benefited from insightful
comments by the anonymous referee. We thank Ken Shen, Alan Calder, and Sam Boos for the useful discussions. We also would like to thank Spencer Caldwell and Broxten Miles for their earlier work on setting up the nova simulation. This work was supported under programs HST-GO-15072, HST-GO-16069, and HST-AR-16638 through the Space Telescope Science Institute, which is operated by the Association of Universities for Research in Astronomy, Inc., under NASA contract NAS5-26555. Support for these programs was provided through a grant from the STScI under NASA contract NAS5-26555.

Software: \texttt{MESA} \citep[][\url{mesa.sourceforge.net}]{Paxton_2011, Paxton_2013, Paxton_2015, Paxton_2018, Paxton_2019}, 
\texttt{GYRE} \citep[][\url{https://gyre.readthedocs.io/en/stable/}]{Townsend_2013, Townsend_2018},
\texttt{Matplotlib} \citep{Hunter_2007}

\nopagebreak
\bibliography{references}{}
\bibliographystyle{aasjournal}

\end{document}